%% file: 0-Manuscript.tex
\documentclass[
 reprint,
 amsmath,amssymb,
 aps,
]{revtex4-2}

\usepackage{graphicx}
\usepackage{dcolumn}
\usepackage{bm}
\usepackage{xcolor}
\usepackage{physics}
\usepackage[top=18mm, bottom=15mm, left=12.5mm, right=12.5mm]{geometry}

\newcommand{\kcalmolAA}{kcal\,mol$^{-1}$\,\AA$^{-2}$}

\newcommand{\fig}[1]{Fig.\,\ref{fig:#1}}
\newcommand{\Fig}[1]{Figure\,\ref{fig:#1}}

\usepackage{amsmath}

\makeatletter
\renewcommand\frontmatter@abstractwidth{\dimexpr\textwidth\relax}
\makeatother

\begin{document}

\title{Electric field of DNA in solution: who is in charge?}

\author{Jonathan G. Hedley}
\affiliation{%
Department of Chemistry, Imperial College London, Molecular Sciences Research Hub, London W12 0BZ, UK
}%
\author{Kush Coshic}
\affiliation{%
\centering Center for Biophysics and Quantitative Biology,
\centering \text{University of Illinois at Urbana-Champaign, Urbana, IL 61801, USA}
}%

\author{Aleksei Aksimentiev}
\email{aksiment@illinois.edu}
\affiliation{%
Department of Physics and Beckman Institute for Advanced Science and Technology, University of Illinois at Urbana-Champaign, Urbana, IL 61801, USA
}%

\author{Alexei A. Kornyshev}%
 \email{a.kornyshev@imperial.ac.uk}
\affiliation{%
Department of Chemistry, Imperial College London, Molecular Sciences Research Hub, London W12 0BZ, UK
}%
\affiliation{
Thomas Young Centre for Theory and Simulation of Materials, Imperial College London, South Kensington Campus, London SW7 2AZ, United Kingdom
}

\begin{abstract}
In solution, DNA, the `most important molecule of life', is a highly charged macromolecule which bears a unit of negative charge on each phosphate of its sugar-phosphate backbone. Although partially compensated by counterions (cations of the solution) adsorbed at or condensed near it, DNA still produces a substantial electric field in its vicinity, which is screened by buffer electrolyte at longer distances from the DNA. This electric field is experienced by any charged or dipolar species approaching and interacting with the DNA. Such field has been explored so far predominantly within the scope of a primitive model of the electrolytic solution, not considering more complicated structural effects of the water solvent. In this paper we investigate the distribution of electric field around DNA using linear response nonlocal electrostatic theory, applied here for helix-specific charge distributions, and compare the predictions of such theory with specially performed fully atomistic large scale molecular dynamics simulations. Both approaches are applied to unravel the role of the structure of water at close distances to and within the grooves of a DNA molecule in the formation of the electric field. As predicted by the theory and reported by the simulations, the main finding of this study is that oscillations in the electrostatic potential distribution are present around DNA, caused by the overscreening effect of structured water. Surprisingly, electrolyte ions at physiological concentrations do not strongly disrupt these oscillations, and rather distribute according to these oscillating patterns, indicating that water structural effects dominate the short-range electrostatics. We also show that (i) structured water adsorbed in the grooves of DNA lead to a positive electrostatic potential core, (ii) the Debye length some $10 \text{ \AA}$ away from the DNA surface is reduced, effectively renormalised by the helical pitch of the DNA molecule, and (iii) Lorentzian contributions to the nonlocal dielectric function of water, effectively reducing the dielectric constant close to the DNA surface, enhances the overall electric field. The impressive agreement between the atomistic simulations and the developed theory substantiates the use of nonlocal electrostatics when considering solvent effects in molecular processes in biology.
\end{abstract}
\maketitle

\vspace{-0.2cm}
\input{1-introduction}
\input{2-nonlocal_electrostatics}
\input{3-theory}
\input{4-theory_results}

\input{5-simulation_methods}
\input{6-simulation_results}

\input{7-conclusions}

\begin{acknowledgements}
This work is supported by the Leverhulme Visiting Professorship grant to AA [VP2-2019-012] which made this work possible. JGH acknowledges support from the Imperial College President's PhD Scholarship. KC and AA acknowledge support from the Human Frontier Science Project (RGP0047/2020) and the National Institute of General Medical Sciences (R01-GM137015). The supercomputer time was provided through the Leadership Resource allocation MCB20012 on Frontera of the Texas Advanced Computing Center and the ACCESS allocation MCA05S028.
\end{acknowledgements}
\vspace{0.5cm}
\textit{Conflicts of interest - } None declared.

\bibliography{bib_DNA}

\input{8-Appendices}
\end{document}

%% file: 1-introduction.tex
\section{Introduction}

A multitude of species interacting with DNA in solution experience its electric field. Indeed, this so-called `most important molecule' is an `electrostatic bomb'. Not actually an `acid', but usually a salt, DNA dissociates in aqueous solution, releasing its cations to the solution and retaining a unit of negative elementary charge on each phosphate of its sugar-phosphate backbone, resulting in two charges per $3.4\text{ \AA}$ vertical rise (or per base pair) of the double helix. The double helix of these negative charges are screened by (i) the Debye cloud of buffer electrolyte, (ii) the DNA's own, released counterions (which are in the minority in an electrolyte solution of physiological concentration $\sim$0.154 M), and (iii) where applicable by counterions (cations) of an added salt that specifically adsorb onto the DNA (see Fig. 1), or those that get condensed just in the narrow layer around DNA~\cite{MANN1978, FRAN1987}. In such an environment the `bomb' is neutralised, but still the electric field, although exponentially decaying into the solution bulk, is substantial in the vicinity of DNA within the range of a few nanometres. This field will act on any charged or polar species, and it thus plays an important role in DNA packing, interaction with proteins, and many other aspects of the vast genetic machinery. 

Studies of the electric field of DNA have a long history, starting shortly after the discovery of DNA structure and function. The first popular model to describe it was the so-called polyelectrolyte model of DNA (for review see~\cite{MANN1978, FRAN1987}). In this model, DNA was considered as a charged cylinder with characteristic DNA radius ($\approx 1$ nm) and mean surface charge density from the two phosphate strands ($\bar{\sigma} \approx 16.3$ $\mu$Ccm$^{-2}$) with the response of the surrounding ions to the presence of such cylinder considered within approximations of various levels of complexity~\cite{GROS2002}. Such models helped to elucidate some features generally in biophysics~\cite{SHAR1990} but were insufficient to unravel effects directly related with the helical structure and symmetry of DNA. 

An attempt to understand the effects of double helical structure on the electric field of a DNA molecule was first made in 1978~\cite{SOUM1978}, but systematic studies of such effects started in the late 90s. It was initially studied in the context of DNA-DNA interaction~\cite{KORN1999}, DNA in dense aggregates~\cite{KORN1998a} and liquid crystals~\cite{KORN2002}, DNA fibres (with reconsideration of structural information that can be extracted from the classical X-ray fibre diffraction patterns)~\cite{KORN2011}, DNA supercoiling~\cite{CORT2011}, and recognition of homologous genes~\cite{KORN2010} (for detailed review see~\cite{KORN2007}). These works have put weight on effects predominantly determined by the helical symmetry of DNA~\cite{KORN1998b} (or violations/distortions of that symmetry~\cite{KORN2009}). Still, to a point, these works all rest on the implicit description of the solvent, describing its dielectric response by a macroscopic dielectric constant, $\varepsilon$, alongside considering the ionic response in a simplified way based on the Debye-Bjerrum approximation ~\cite{KORN2007}. A series of publications were devoted to account for the nonlinear response of the ionic subsystem, based on concepts of Wigner-crystal formation~\cite{SHKL1999} and strong ionic correlations~\cite{GROS2002}, but they did not consider the helicity of the DNA charge distribution, and again, neither went beyond a macroscopic description of the dielectric response of the surrounding water. 

It has been, however, known for several decades that macroscopic dielectric response is insufficient in the description of electrostatics in water. For example, let us simply consider a simple single ion: submerging this in water will create a solvation sphere of bound solvent molecules, putatively resulting in an effective, reduced dielectric screening close to the ion. As polarisation fluctuations in polar media are correlated in space (in the case of water, by its hydrogen bond network), this effect will persist over a certain characteristic length intrinsic to the solvent, and the effective dielectric constant will return to its bulk value only far away from the ion, at distances longer than this characteristic length. Given that this dielectric constant is effectively distance dependent, it is natural to refer to the language of nonlocal electrostatics, where water polarisation correlations are conveniently described by the wave-number dependent dielectric function, $\varepsilon(k)$; in the linear response approximation such an approach provides a framework within which we can analyse the effect of water structure and dielectric response to any charged object. 

Simple models of the dielectric response of water interpolate smoothly between macroscopic ($\varepsilon\approx 80$) and high-frequency ($\varepsilon_*\approx 3-5$) dielectric constants, which leads to additional exponentially decaying contributions associated with the water structure in the electrostatic potential distribution near a charged or polar species embedded in water. As such it will manifest itself in the potential distribution near an electrode, in hydration forces between charged or polarised surfaces, etc. Exponentially varying ‘structural’ contributions to the forces between objects at the nanoscale have been measured in many biologically relevant systems from lipid membranes to DNA and proteins, as well as between surfaces relevant in electrochemistry. 

Water is, however, more complex than this. Detailed analysis of its dielectric response shows so-called `overscreening’ effects in addition to these exponential correlations. This effect means that in the first molecular layer around a charged species, the amount of bound countercharge is larger than that on the species; that excess is overcompensated in the next layer, and off it goes until the macroscopic limit of screening is reached. This manifests itself in a peak in the wave-number dependent Fourier component of the response function, $\chi(k)=\varepsilon_*^{-1}-\varepsilon(k)^{-1}$. This implies that there exists a region where the dielectric function $\varepsilon(k)$ can be negative, and will lead to oscillations in the electrostatic potential around the species or in hydration forces. Such oscillations have been observed by Israelachvili and Pashley in 1983~\cite{ISRA1983} in their force measurements with surface force apparatus (SFA) between atomically flat mica surfaces. Such oscillations have also been seen even in the earliest computer simulations~\cite{PHIL1995a, PHIL1995b, SPOH1998, SPOH1999, DIMI2000}. This begs the question; why were these oscillations not seen previously in force experiments between lipid membranes or even differently prepared mica surfaces? In their paper, Israelachvili and Pashley alluded to the roughness of the surfaces in question; structural and thermal fluctuations can disrupt the water structure, leading to oscillation dysphasia. Such a reasoning is logical; recently this conjecture has been substantiated by a systematic theoretical analysis~\cite{HEDL2023}. 
\begin{figure}[t!]
\centering
\includegraphics[width=\linewidth]{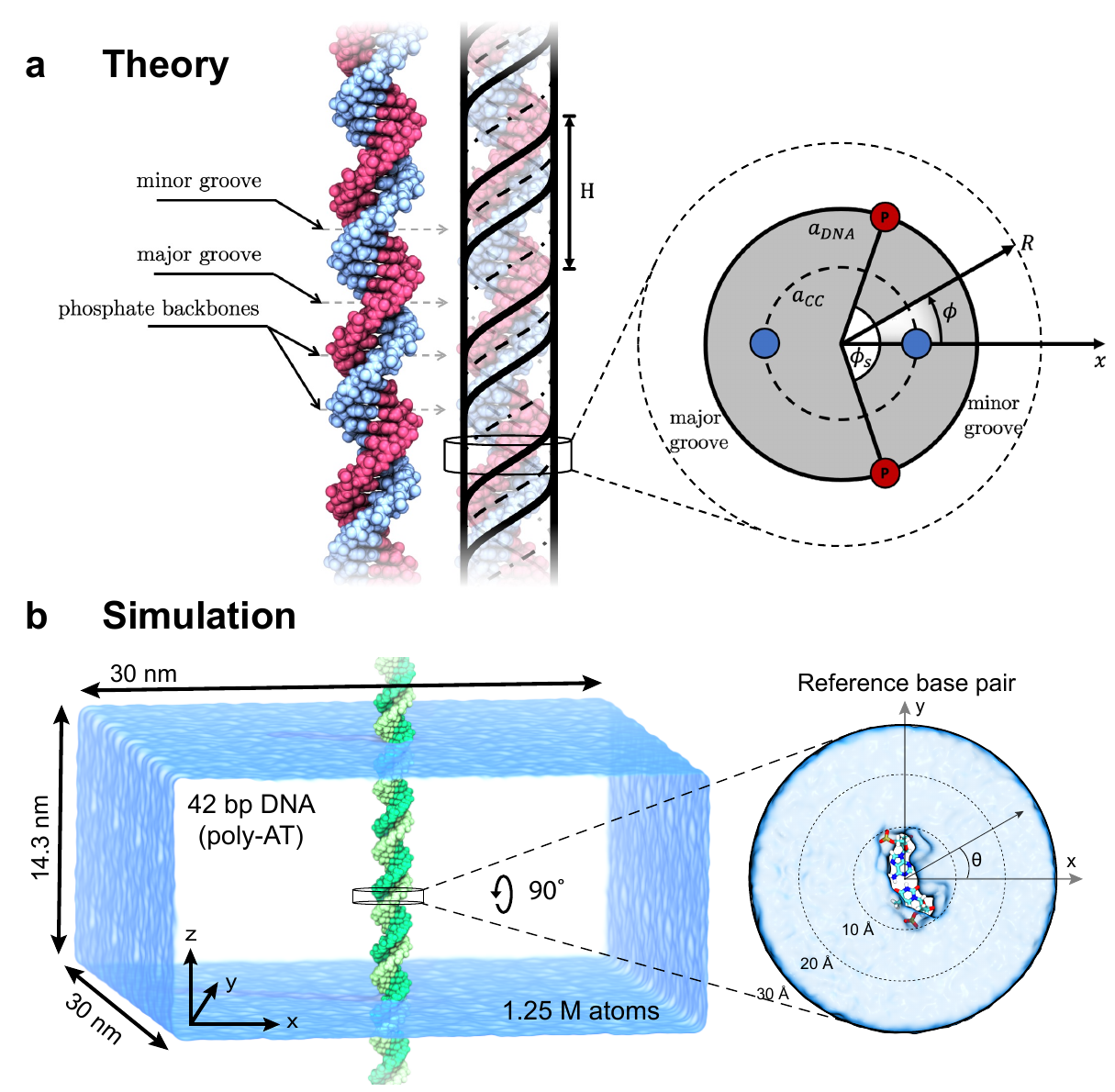}
\caption{\textbf{Comparison between theory and simulation models for a DNA molecule}. \textbf{(a)} Illustration of DNA surface charge pattern, consisting of negatively charged helical lines of phosphates (solid) and positively charged counterions adsorbed in the major (dot-dashed) and minor (dashed) grooves. Such an ideal helical surface charge pattern corresponds closely to poly-AT DNA, which has helical pitch $H\approx34$ \AA. On the right of (a), we show a cross-section of the DNA molecule at $z=0$, displaying the polar coordinate system $(R, \phi)$ we use in this work. Phosphates (red circles) sit at a radius of $a_{\text{DNA}}\approx 10$ \AA, where $\phi_s$ is the azimuthal width of the minor groove. Specifically adsorbed counterions (blue circles) sit within the DNA grooves at a radius of $a_{\text{CC}}\approx 5-7$ \AA. Not shown in the figure are the helical lines of structured water (the `water spine') which also sit in the grooves at a radius of $a_{\text{W}}\approx 4-5$ \AA. \textbf{(b)} Illustration of the simulation system, consisting a 42 base pair DNA (poly-AT) submerged in solvent (semi-transparent surface). Each strand of the duplex is covalently linked to itself across the periodic boundary, making the DNA effectively infinite. A cylindrical volume of radius 3 nm, centered around the central base pair of the duplex, is taken as a reference. The inset defines a coordinate system perpendicular to the DNA’s helical axis.}
\label{fig:Figure 1}
\end{figure}
These oscillations are currently receiving significant attention and are being measured more and more frequently in a number of sophisticated frequency modulated atomic force microscopy (FM-AFM) experiments where it is possible to bypass the effect of roughness under the so-called `solvent-tip approximation’~\cite{WATK2013, KLAA2022}. However, the exact consequence and relevance of such oscillating features on phenomena such as ion adsorption and double layer structure is still under debate~\cite{ADVI2024}. For the case of DNA, its large relative size compared to the solvent molecules in combination with its double helical structure and thermal fluctuations would lead one to naively believe that all these oscillations would be smeared out. However, recent FM-AFM experiments measuring the hydration structure of DNA has shown that we do indeed see signatures of oscillations in the force patterns~\cite{KUCH2018}. Such observation is consistent with the idea of a ‘DNA water spine’; a stabilising chiral superstructure that has also been experimentally measured in X-Ray diffraction experiments~\cite{MCDE2017}. Hence, when considering the complex electric environment surrounding a single DNA molecule, we must resort to a more sophisticated model which can incorporate these oscillating features. 

More recently, the all-atom molecular dynamics (MD) simulation method has become a ubiquitous tool for accurate characterisation of biomolecular systems~\cite{KARP2002A}. The application of this method to the study of DNA systems has been challenging because the high charge of the DNA molecules combined with physiological salt concentrations requires larger systems and long simulation times to fully equilibrate the ion atmosphere. Nevertheless, early work has shown that a fully atomistic MD model can reproduce DNA electrostatics inferred from continuum simulations and DNA supercoiling measurements~\cite{MAFF2010B}. Soon after, however, the standard parameterization of ion--DNA interactions was found to be inadequate to reproduce experimentally measured DNA--DNA forces~\cite{YOO2012a}. The force field model was then refined by introducing surgical corrections to cation--DNA phosphate interactions, producing a molecular force field model capable of quantitative reproduction of the DNA array data~\cite{RAU1984,YOO2016B} and competitive ion binding experiments~\cite{BAI2007,YOO2012b}. The model was then used to \textit{predict} the effect of DNA methylation on DNA--DNA forces~\cite{YOO2016C}, suggesting a physical mechanism for guiding DNA condensation into microcompartments according to the DNA sequence~\cite{KANG2018}.

With these computational advances in mind, the question remains; how much will things change when we abandon the previously established macroscopic models of DNA electrostatics and try to account for water structure around DNA? What effect does the water structure have on the distribution of ions around DNA, and what does it all matter for the `fine structure' of the electric field created by a DNA molecule in solution? Is a theoretical approach to this problem unrealistic, and are these results accessible only through all-atom MD simulations?

This article is a first attempt to answer some of these questions. We will incorporate a dielectric function inspired by a field-theoretical account of water structure into the dielectric response of the electrolyte to the helical charge distribution representing DNA. Calculating the electrostatic potential, electric field and charge density, we will analyse results both in electrolyte solution and in a hypothetical `pure water' case to understand the nature of the electrostatics of the system. With those tasks achieved, we will perform fully atomistic molecular dynamics simulations of DNA, water, and electrolyte ions, the approach which in principle has its own extensive history~\cite{JAYA1996, MAFF2014} and compare the results with those of the analytical theory. 

Previewing, we can conclude that the results of these two approaches appear in harmony with each other, which is especially important in the absence of direct experimental determination of the distribution of electric field around DNA. Specifically, the theoretical approach to the description of the dielectric response of the water solvent is based on (and limited by!) linear nonlocal electrostatics ~\cite{KORN1981, KORN1980}, and thus the importance of the verification of its predictions by fully atomistic computer simulations is obvious.

The structure of this article is as follows. Before presenting the analytical theory for the calculation of electric field about DNA, we first provide an `elevator pitch' of the principles of nonlocal electrostatics, illustrating and discussing its predictions for species much simpler than DNA. We then proceed to the basics of the analytical theory of the electric field of DNA in solution, and present its results and predictions. Following this, we describe the set up and parameters of the simulations performed in this study. Then, having covered the foundations of both the theoretical and computational methods, we present the results of the simulations and compare them to the predictions of the theory, discussing the consequences and findings of this in-depth study into the electrostatics of this `most important molecule'.

%% file: 2-nonlocal_electrostatics.tex
\section{Nonlocal electrostatics:\\ A Birds-Eye view}

In this work, we use the language of nonlocal electrostatics to understand the effect of structured water surrounding a charged object. To give the reader more context as to the applicability and validity of this generalisation of classical electrostatics, we first provide a whistle-stop tour of the main concepts of the approach below, applying it to one of the simplest models of an ion, the Born sphere. We do this to show the degree of complexity required when more complicated features, such as oscillations in water structure, are considered. Note that Gaussian units are used throughout the article in all mathematical formalism.

\subsection{Basic Equations of Nonlocal Electrostatics}

When we consider spatially correlated media, the displacement field, $\mathbf{D}$ and polarisation density field, $\mathbf{P}$ at a point $\mathbf{r}$ are not simply proportional to the electric field, $\mathbf{E}$, as in the constitutive relations of classical electrostatics, i.e. $\mathbf{D}(\mathbf{r})=\varepsilon\mathbf{E}(\mathbf{r})$ and $\mathbf{P}(\mathbf{r})=\chi\mathbf{E}(\mathbf{r})$, where $\varepsilon$ and $\chi$ are respectively the dielectric constant and dielectric susceptibility of the medium. But generally, $\mathbf{D}(\mathbf{r})$ and $\mathbf{P}(\mathbf{r})$ must depend on the electric field in the surrounding space of that point because there are spatial correlations in the system; in other words, the orientation of one dipole in an electric field in space depends on how its surrounding dipoles are oriented, i.e. it depends on the value of the electric field in the volume around the point $\mathbf{r}$ extending to the range of spatial correlation of polarisation fluctuations. Hence, the central idea of nonlocal electrostatics is the generalisation of these constitutive relations of classical electrostatics into the nonlocal form:
\begin{align}
    \mathbf{D}_{\alpha}(\mathbf{r}) = \sum_{\beta}\int d\mathbf{r}' \varepsilon_{\alpha\beta}(\mathbf{r}-\mathbf{r}')E_{\alpha\beta}(\mathbf{r}'),
\end{align}
where the subscripts denote Cartesian components, $\alpha,\beta=x,y,z$. The above-mentioned correlations manifest themselves in the kernel of the relation, $\varepsilon_{\alpha\beta}(\mathbf{r}-\mathbf{r}')$, the so-called nonlocal dielectric tensor. In the limit of macroscopic electrostatics, $\varepsilon_{\alpha\beta}(\mathbf{r}-\mathbf{r'})=\varepsilon \delta_{\alpha\beta}\delta(\mathbf{r}-\mathbf{r'})$, where $\delta_{\alpha\beta}$ is the Kronecker delta, and $\delta(\mathbf{r}-\mathbf{r}')$ is the Dirac delta function. Such form reduces this general nonlocal relation to the local expression. In homogeneous and isotropic media, it is convenient to instead consider the Fourier transform of this tensor, $\varepsilon_{\alpha\beta}(\mathbf{k})$, or more precisely its longitudinal component:
\begin{align}
    \varepsilon_{||}(k)=\sum_{\alpha\beta}\frac{k_\alpha k_\beta}{k^2}\varepsilon_{\alpha\beta}(\mathbf{k}),
\end{align}
where $k=|\mathbf{k}|$, which we will express as $\varepsilon(k)$ below for brevity.

The potential of the field produced by an arbitrary charge distribution in a uniform medium is determined by Gauss' law, $\nabla\cdot\mathbf{D}=4\pi\rho_{ext}(\mathbf{r})$. After the substitution of Eq. 1 with $\mathbf{E}(\mathbf{r})=-\nabla\varphi(\mathbf{r})$, one has:
\begin{align}
    \sum_{\alpha\beta}\pdv{}{r_\alpha}\int d\mathbf{r}'\varepsilon_{\alpha\beta}(\mathbf{r}-\mathbf{r}')\pdv{}{r_\beta}\varphi(\mathbf{r}') = -4\pi\rho_{ext}(\mathbf{r}).
\end{align}
We can easily resolve this equation with respect to the potential $\varphi$. With the following definitions of the forwards and inverse Fourier transforms of any function $f(\mathbf{r})$
\begin{align}
    \tilde{f}(\mathbf{k})=\int d\mathbf{r}\hspace{1mm}f(\mathbf{r})e^{-i\mathbf{k}\cdot\mathbf{r}}\quad ; \quad  f(\mathbf{r})=\frac{1}{(2\pi)^3}\int d\mathbf{k}\hspace{1mm}\tilde{f}(\mathbf{k})e^{i\mathbf{k}\cdot\mathbf{r}}
\end{align}
as well as Eq. 2, we obtain an expression for the potential produced by a rigid distribution of external charges immersed in a nonlocal solvent,
\begin{align}
    \tilde{\varphi}(\mathbf{k}) = \frac{4\pi}{k^2\varepsilon(k)}\tilde{\rho}_{\text{ext}}(\mathbf{k}).
\end{align}
From this expression, the first step we need is to determine the form of $\varepsilon(k)$ for the solvent. In the past, the `Lorentzian' form of the dielectric function which interpolates between the limiting behaviour at small and large $k$ was widely used. Such an expression for $\varepsilon(k)$ describes purely exponentially decaying correlations in the system, which was able to rationalise many different experimental observations in a number of electrochemical systems~\cite{KORN1981, KORN1980}. However, this so-called `Lorentzian model' fails to capture the full complexity of the system; it does not describe the overscreening mode in the dielectric response functions ubiquitous to polar liquids, where $\varepsilon(k)<0$. We can write a general expression for the dielectric function as follows:
\begin{align}
    \varepsilon(k) = \frac{1}{\dfrac{1}{\varepsilon_*}-\left(\dfrac{1}{\varepsilon_*}-\dfrac{1}{\varepsilon}\right)\tilde{F}(k)}
\end{align}
where $\varepsilon_*\approx 3-5$ is the short-range dielectric constant. Within the Lorentzian model, $\tilde{F}(k)=1/(1+\Lambda^2k^2)$, where $\Lambda$ is the correlation length of the polarisation fluctuations in the liquid. This model does not consider finer effects in the $k$-spectrum, leading to overscreening. Taking inspiration from a dielectric function derived from an extended phenomonelogical Landau-Ginzburg expansion in the polarisation density of water (see Ref.~\cite{HEDL2023} for details), we use the following expression, first proposed in Ref.~\cite{LEVY2020}:
\begin{align}
    \tilde{F}(k) = \frac{\gamma}{1+\Lambda_1^2k^2} + \frac{(1-\gamma)(1+\Lambda_2^2Q^2)^2}{(1+(k\Lambda_2-Q\Lambda_2)^2)(1+(k\Lambda_2+Q\Lambda_2)^2)}
\end{align}
Such form for the dielectric function accounts for a number of fine-structure effects in water. This is particularly clear if we examine the poles of Eq.(7). There are six roots of the denominators, two of which are imaginary, located at $k=\pm i/\Lambda_1$, and the rest being complex, located at $k=\pm Q \pm i/\Lambda_2$, where $Q, \Lambda_1$ and $\Lambda_2$ are related to the different correlation lengths in the system, and we take their values from fits to simulated bulk water dielectric response function~\cite{KORN1996, HEDL2023}. The two imaginary poles describe exponential (a.k.a. Lorentzian) correlations, with characteristic decay length $\Lambda_{1}\approx 3.67$ \AA. The complex poles describe decaying oscillating correlations, with characteristic decay length $\Lambda_{2}\approx 1.77$ \AA, and oscillation period $\Lambda_{o}=2\pi/Q \approx 2.13$ \AA. The partitioning coefficient $\gamma$ determines the relative strength of these contributions to the overall dielectric response. Here we have chosen parameters such that $\varepsilon(k)$ reproduces the response function for TIP3P water, the model used in the following DNA simulations (see \textbf{\textit{Simulation Methods}}). It is important to note that when comparing the results of the theory against simulations performed with such rigid bond models which neglect internal degrees of freedom of the water molecule, such as electronic polarisability and bond vibrations, we can simply set $\varepsilon_* = 1$. However, when more detailed models of the solvent are used in simulation, we must consider more carefully the value of this $\varepsilon_*$, as well as including the varying spatial dispersion of these internal degrees of freedom.  

\begin{figure*}[t!]
    \centering
    \includegraphics[width=\linewidth]{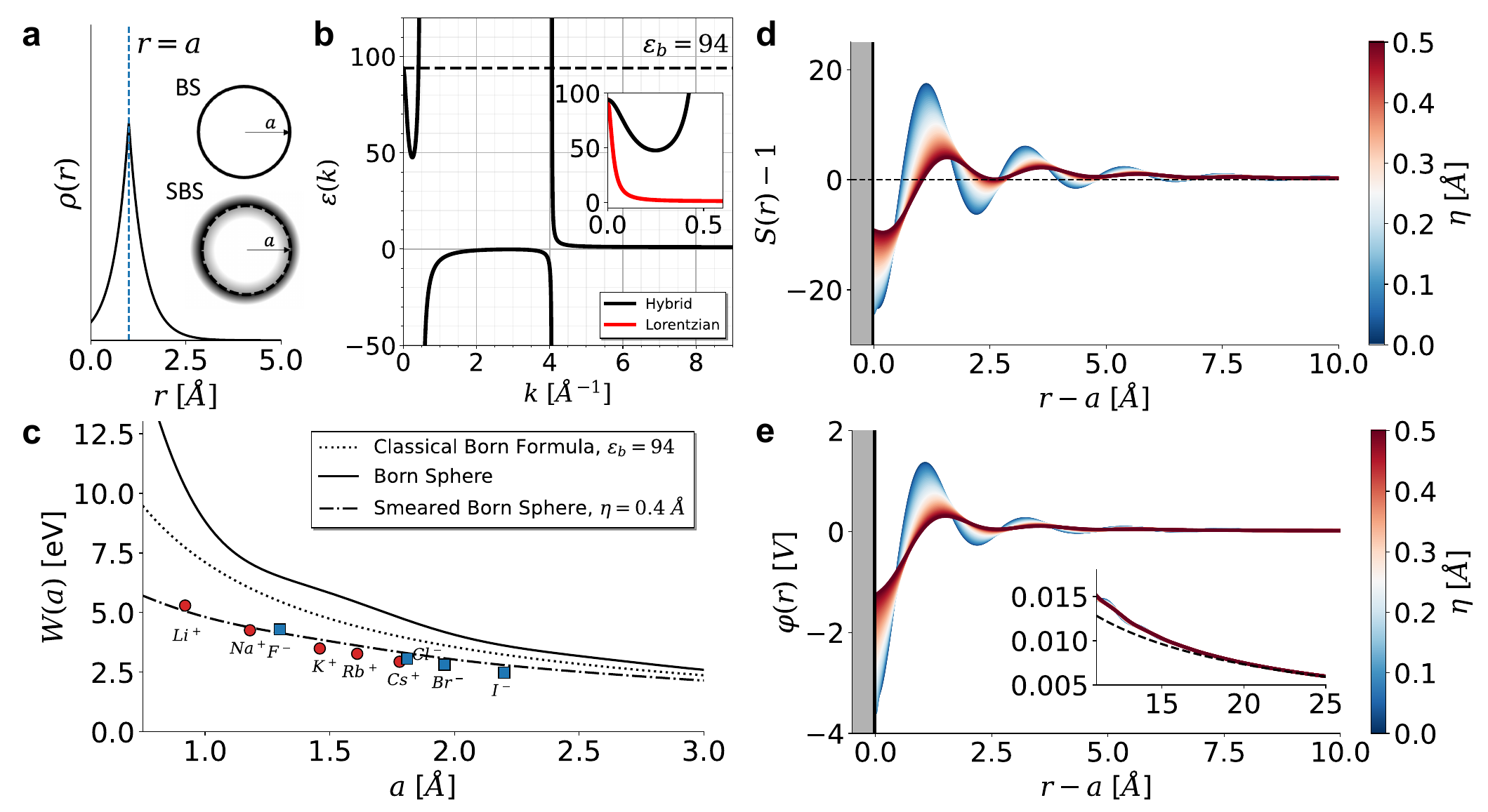}
    \caption{\textbf{Nonlocal Electrostatics results for a smeared Born sphere with hybrid dielectric function parameterised to TIP3P:} 
    \textbf{(a)} Model charge distributions for $a=1 \text{ \AA}$ and $\eta = 0.4 \text{ \AA}$.  Vertical dotted blue line represents the BS model, and the solid black line represents the SBS model. \textbf{(b)} Hybrid model for the dielectric function of water, defined by Eqs.(6) and (7), parameterised for TIP3P water, with $\Lambda_1=3.67 \text{ \AA}$, $\Lambda_2=1.77\text{ \AA}$, $\Lambda_o=2\pi/Q=2.13 \text{ \AA}$, $\gamma=0.05$ and $\varepsilon_*=1$. At $k=0$, $\varepsilon(k=0) = \varepsilon = 94$ for TIP3P water, as shown by the horizontal dotted line. Inset shows comparison between the hybrid and Lorentzian models in the small-$k$ region. Figures \textbf{(c)}, \textbf{(d)} and \textbf{(e)} all show quantities calculated using this model of $\varepsilon(k)$. \textbf{(c)} Theoretical calculations against experimental measurements for hydration energy against ion radius. Experimental data for ions taken from Ref.~\cite{RAND1956}. We clearly see that the classical Born formula and the hard BS model overestimate the hydration energy significantly. However, when smeared, the theoretical predictions match experiment almost exactly. \textbf{(d)} and \textbf{(e)} show the screening function and the electrostatic potential distribution respectively, and how they vary with the smearing parameter, $\eta$. Inset of \textbf{(e)} shows convergence with the classical Coulomb law at distances larger than $\sim 20$ \AA, where the effect of polarisation correlations disappears.}
    \label{fig:2-NEresults}
\end{figure*}
    
\subsection{Solvent response to the electric field of simple ions}
Rather than diving headfirst into the more complicated case of DNA, we will begin by studying a single ion. The simplest model one can adopt for this is the Born sphere model (BS), which has been widely used in many electrostatic calculations in the past. Here, the ion charge $Q$ is localised on a sphere of radius $a$, such that 
\begin{align}
    \rho_{\text{BS}}(\mathbf{r}) = \frac{Q}{4\pi a^2}\delta(|\mathbf{r}|-a).
\end{align}
Such a model is quite crude; simple arguments from quantum mechanics tell us that a hard sphere of charge is a fairly poor model of an ion. Firstly, given the presence of directional orbitals surrounding the ion, the charge distribution is generally not isotropic. Secondly, electrons are not localised on an infinitely thin sphere, but rather can be thought of as `smeared' along the radial direction. We therefore consider a smeared Born sphere model (SBS) introduced in Ref.~\cite{KORN1996}. This is one of the simplest modifications of the Born model, neglecting anisotropy in the charge distribution.  The charge distribution of such a sphere is given by
\begin{align}
    \rho_{\text{SBS}}(\mathbf{r}) = \frac{QN_{\text{SBS}}}{4\pi}e^{-||\mathbf{r}|-a|/\eta}.
\end{align}
Here, $a$ is the position of maximum charge density (effective ion radius) and $\eta$ is the smearing parameter. We determine the normalisation factor, $N_{SBS}$, from the condition that $\int_V\rho(\mathbf{r})d\mathbf{r} = q$, yielding
\begin{align}
    N_{\text{SBS}} = \frac{1}{2\eta(a^2+\eta^2(2-e^{-a/\eta}))}.
\end{align}
The classical Born model is a limiting case of this smeared model, as the smearing parameter $\eta\to 0$.
The Fourier transforms of these charge distribution models are given by
\begin{align}
    \tilde{\rho}_{\text{BS}}(\mathbf{k})=Q\frac{\sin(ka)}{ka},
\end{align}
\begin{align}
    \tilde{\rho}_{\text{SBS}}(\mathbf{k})=2QN_{\text{SBS}}\left\{\frac{\eta a \sin(ka)}{k(1+\eta^2k^2)}+\frac{\eta^3(2\cos(ka)-e^{-a/\eta})}{(1+\eta^2k^2)^2}\right\}
\end{align}
where we clearly see that Eq. 13 reduces to Eq. 12 in the case where $\eta\to0$. 
\vspace{-0.3cm}
\subsubsection{Electrostatic Potential} To calculate the electrostatic potential around a spherically symmetric ion, we can simply take the spherical 3D inverse Fourier transform of Eq.(5), yielding the nonlocal electrostatic formulation,
\begin{align}
    \varphi(r)=\frac{2}{\pi}\int_0^\infty \frac{dk}{\varepsilon(k)}\frac{\sin(kr)}{kr}\tilde{\rho}_{\text{ext}}(k).
\end{align}
From this, we can also define the screening function $S(r) = \varepsilon r\varphi(r)$, which yields the deviation from Coulomb's law for the electrostatic potential due to a monovalent ion. 

Limitations of this equation are as follows; (i) linear response of the dielectric medium to the electric charge, (ii) neglect of excluded volume of the charged object itself (the so-called `embedded charge approximation'). These limitations were both analysed in Ref.~\cite{FEDO2007}, where it was shown that very close (within one molecular diameter) to the ion, both effects lead to substantial differences in the distribution of electrostatic potential. Hence, we must acknowledge that the results we obtain through this nonlocal formalism within the range of one molecular diameter (here $\sim 2.5$ $\text{\AA }$  for water) should be considered with a pinch of salt. As can be seen in Figs.  2(d) and (e) to follow, these limitations manifest as a possible artefact, where we see inversion in the sign of the potential and screening factor very close to the ion surface. 
\vspace{-0.2cm}
\subsubsection{Free enthalpy of solvation} The free enthalpy of solvation, which we term here the hydration energy, is generally given as the difference between the electrostatic energy in a vacuum and in the solvent:
\begin{align}
    W=\frac{1}{2}\int_V d\mathbf{r}\left[\varphi_0(\mathbf{r})-\varphi(\mathbf{r})\right]\rho_{\text{ext}}(\mathbf{r}).
\end{align}
Using Eq.(13) above for the electrostatic potential, we obtain~\cite{KORN1981}:
\begin{align}
    W(a)=\frac{1}{\pi}\int_0^\infty dk \hspace{1mm}\tilde{\rho}_{\text{ext}}(k; a)^2\left[1-\frac{1}{\varepsilon(k)}\right].
\end{align}
It is important to note that such expressions for the hydration energy assume that the solvent penetrates inside the charge distribution (a.k.a. the embedded charge distribution approximation~\cite{KORN1981, KORN1996}); such an assumption works reasonably well as an interpolation: when the ion radius is small and nonlocal effects are the largest, the amount of solvent ‘within the ion’ is negligible, whereas for larger ions the nonlocal effect diminishes and the assumption that the solvent sits inside the ion bears no importance, as in the classical Born formula. Note that in the limit where $\varepsilon(k)=\varepsilon=\text{const.}$, and $\Tilde{\rho}(k;a)\to\tilde{\rho}_{BS}(\mathbf{k})$, Eq.(15) reduces to the classical Born formula:
\begin{align}
    W(a) = \frac{1}{2a}\left(1-\frac{1}{\varepsilon}\right).
\end{align}
Plotting the hydration energy against ion size in Fig. \ref{fig:2-NEresults}(c), we see that for large ions, the nonlocal expression for the Born sphere approaches the classical expression asymptotically. It is well known that that the classical Born formula overestimates the hydration energies for ions (Fig.2c). This arises due to the approximation that $\varepsilon(k)\equiv\varepsilon$ in all space, including in the vicinity of the ion, not accounting for nonlocal effects. When we calculate this hydration energy for a Born sphere in a nonlocal medium within the Lorentzian approximation, the value of the hydration energy successfully reduces down to experimental values~\cite{KORN1981}. When, however, we try to use more sophisticated models which account for overscreening, the overestimation worsens (Fig.2c). This was shown to be an artefact of the BS model~\cite{KORN1996}. Calculating the hydration energy for the more realistic SBS model allows us to match the experimental results much more closely by simply smearing the charge slightly~\cite{KORN1996}, where we have set $\eta=0.4 \text{ \AA}$ for all ions. Of course, each ion will have its own characteristic $\eta$, and so it is not accurate to simply apply a constant value for all ionic radii. However, this example just shows the absolute importance of introducing smearing in the charge distribution when the dielectric function accounts for overscreening oscillations.
\vspace{-0.2cm}
\subsection{Interpolation Approximation for Electrolytes}

Recently, a phenomenological model for the dielectric function of pure water was proposed~\cite{HEDL2023}, however the approach to include electrolyte ions there was only valid for small concentrations so as to not violate the Dolgov-Kirzhnits-Maksimov (DKM) constraint~\cite{DOLG1981} on the full electrolyte dielectric function, $\varepsilon_c(k)$. This constraint means that $\varepsilon_c(k)$ cannot enter the regime of $0<\varepsilon_c(k)\leq 1$ at any concentration of electrolyte. However, within the following approximation first used in Ref.~\cite{LEVY2020}, we can extend the dielectric function to account for both the solvent molecules and the ions present without worry of violating this DKM law.  

Let us remind ourselves of the limiting behaviour of the dielectric function; in the long wavelength limit (small $k$), we recover macroscopic behaviour, i.e. $\varepsilon(k)\to\varepsilon$, whereas the short wavelength limit (large $k$) probes the short-range correlations. For example, when we consider pure water, the wave-numbers $k\sim 2\pi/d$, where $d$ is the diameter of the water molecule, characterise the molecular packing effects in the solvent. This is the origin of the oscillation period $\Lambda_o (=d)$, obtained from the roots of $\varepsilon(k)$. For much larger $k$, $\varepsilon(k)$ will approach the short-range dielectric constant due to the electronic/infrared polarisability of the molecules, $\varepsilon_*$. 

Analysis of the linear Poisson-Boltzmann equation for a binary monovalent solution yields the dielectric response function of ionic solutions in the long wavelength limit (small $k$):
\begin{align}
    \varepsilon_{c}(k)=\varepsilon\left[ 1+\frac{\kappa^2}{k^2} \right]
\end{align}
where $\kappa^{-1}=\lambda_D$ is the Debye screening length. The divergence at small wavenumbers corresponds to the screening of the potential at distances larger than the Debye length. At smaller distances, the screening effect is negligible, and the dielectric response is only influenced by the water. This will remain true even if we consider a more complicated expression for the water dielectric response, rather than $\varepsilon$. Hence, we can write a simple interpolated formula for the dielectric response by replacing $\varepsilon$ with the full $\varepsilon(k)$:
\begin{align}
    \varepsilon_c(k) = \varepsilon(k)\left[1+\frac{\kappa^2}{k^2}\right]
\end{align}
where $\varepsilon(k)$ is the pure water dielectric function that we can approximate by, e.g., Eqs. 6 and 7. Checking the limiting behaviour of the expression, for long wavelengths, $\varepsilon(k)\to \varepsilon$, and we recover Eq. 17. In the short wavelength limit, the ionic contribution is neglected and we recover the pure water response. By design, this interpolated formula is expected to work well if there is a separation of length scales: the Debye length is larger than the characteristic correlation lengths in the solvent, and for sure much larger than the size of the solvent molecules. It is this interpolated formula for the electrolyte dielectric function that we will use in this paper in the Poisson Equation to solve for the potential around a DNA molecule. 

Of course, writing a general dielectric function coupling the solvent to the ionic response is not an easy task. The approach outlined above, in the simplest way, is equivalent to replacing the inverse Debye length $\kappa$ with some wavenumber-dependent $\tilde{\kappa}(k)$, a sophistication that has received a lot of attention recently in the attempts to combine the dielectric response of the solvent to the electrolyte ions~\cite{KJEL2019, KJEL2016, VARE1998, BUDK2020}. However, in these works, the method by which we obtain this $\tilde{\kappa}(k)$ is not straightforward, and results in no simple formula. Hence, we use Eq. (18) here as a first approximation, keeping in mind its limitations. 

In view of a number of the approximations used, such as the one inherent to Eq.(18), as well as the model for the dielectric function of water (although fitted to independent molecular dynamics simulations), we will systematically compare the predictions of the theoretical results with atomistic molecular dynamics simulations, which is the cornerstone of this paper.

%% file: 3-theory.tex
\vspace{-0.4cm}
\section{Theory \& Model}
\label{theory}
\subsection{Basic Equations}

Let us now consider one infinitely long, cylindrical molecule with an arbitrary surface charge distribution. Here we assume that an aqueous solution fills all the space, and the fixed charges are immersed in it. Given that water is able to penetrate within the grooves of DNA, this is not too crude an assumption to make. It should be noted however that this water has structure, resulting in a net polarisation density, and therefore must be included in the model for the surface charge distribution as a `bound charge' contribution (see sub-section on \textit{\textbf{Bound Charge Distribution}}). 

Using the approximation detailed above for the dielectric function of electrolytes, the Fourier transform of the electrostatic potential $\varphi(\mathbf{r})$ created by any embedded charge distribution of volume charge density $\rho_{ext}(\mathbf{r})$ is given by:
\begin{align}
    \tilde{\varphi}(\mathbf{k}) =\frac{4\pi}{(k^2+\kappa^2)\varepsilon(k)}\tilde{\rho}_{ext}(\mathbf{k}).
\end{align}
where $\tilde{\rho}_{ext}(\mathbf{k})$ is the Fourier transform of $\rho_{ext}(\mathbf{r})$. As we are dealing with the surface charge density on a cylindrical surface, it is convenient for us to describe the external charge density in the molecular frame in a cylindrical coordinate system, $(z,\phi, r)$, associated with the molecular axis (see Fig.3(a)). There are multiple contributions to the external charge distribution (see below) - to keep our formalism as general as possible, we can write the charge density of a given contribution $\nu$ as an arbitrary surface charge distribution, placed at a given radius $a_\nu$, such that:
\begin{align}
    \rho_{\nu}(z,\phi, r) = \sigma_\nu(z,\phi)\delta(r-a_\nu)
\end{align}
It is then convenient for us to express it in the form:
\begin{align}
    \rho_{\nu}(\mathbf{r}) = \frac{1}{(2\pi)^2}\sum_{n=-\infty}^{\infty} \int_{-\infty}^{\infty}dq\hspace{1mm}e^{iqz}e^{in\phi_{\mathbf{R}}}\tilde{\rho}_{\nu}(q,n,r).
\end{align}
so that
\begin{align}
    \tilde{\rho}_{\nu}(\mathbf{k}) =  a_\nu\sum_{n=-\infty}^{\infty} i^n\Tilde{\sigma}_\nu(q,n)J_n(Ka_\nu)e^{-in\phi_{\mathbf{K}}}.
\end{align}
where $\tilde{\sigma}_\nu(q,n)$ is the Fourier transform of $\sigma_\nu(z,\phi)$, defined as
\begin{align}
    \tilde{\sigma}(q,n) = \int_0^{2\pi}d\phi \int_{-\infty}^\infty dz\hspace{1mm}e^{-iqz}e^{-in\phi}\sigma(z,\phi),
\end{align}
and $J_n(x)$ is the $n$-th order Bessel function of the first kind. Plugging Eq. (22) into (19), and performing the inverse Fourier transform, we find the expression for the potential in real space:
\begin{align}
    &\varphi_\nu(\mathbf{r}) = \frac{a_\nu}{\pi}\sum_{n=-\infty}^{\infty}\int_{-\infty}^{\infty} dq \int_0^\infty KdK\nonumber\\
    &\hspace{2cm}\times e^{in\phi}e^{iqz}\frac{J_n(KR)J_n(Ka_\nu)}{\varepsilon(\sqrt{K^2+q^2})(K^2+q^2+\kappa^2)}\tilde{\sigma}_\nu(q,n)
\end{align}
which will give us a result for a given $\tilde{\sigma}_\nu(q,n)$ and $\varepsilon(k)$. In previous approaches to this problem, this expression has been extended to include the presence of low dielectric cylindrical cores, such that the charge distribution sits at the inner-core/water interface~\cite{KORN1997}. Considering that water does sit in both the major and minor grooves and because we also consider spatial dispersion of the solvent, we may neglect such effects, accounting for which would have had very much complicated the theory. We will therefore remain with this picture of a charge distribution immersed in solution.

\subsection{Surface charge distribution model}
The expression for the potential in Eq. (24) is valid for any charge distributed over concentric cylindrical surfaces of radii $a_
\nu$. Previous formulations of the theory considered simple examples of infinitely thin, continuous, homogeneously charged helical lines (one or several), and a homogeneously smeared counter-charge both located at the cylinder/water interface~\cite{KORN1997}, as well as smeared lines~\cite{KORN1999} or discrete charge arrays~\cite{KORN1998b}. Given that we will introduce a dielectric function that includes the overresponding behaviour of water, these infinitesimally narrow lines of charge will introduce very large spatial oscillations of electrostatic potential into the system, as reasoned above when considering even just single ions. We therefore need to take into account more complex effects associated with the finite size of the charged groups, on- and off- strand fluctuations around their regular positions on the helical strands, and inhomogeneous distributions of adsorbed counter-charges, which will smear out these resonant overscreening effects and provide more realistic results.  

\subsubsection{Backbone Charge Distribution} Here, we consider the DNA molecules to have ideal helical symmetry, which would imply that the relation
\begin{align}
    \sigma(z,\phi)=\sigma(z+z', \phi+gz')
\end{align}
must be satisfied for any $z'$, where $g=2\pi/H$, and $H$ is the helical pitch (for B-DNA, $H=34\text{ \AA}$). We see there that there is an equivalency between $z$ and $\phi$: a change in $z$ by $z'$ is equivalent to rotating the molecule through an angle of $gz'$. We therefore approximate the density of fixed surface charges, intrinsic to the helical molecules, by the following general expression, valid for any helical molecule:
\begin{align}
    &\sigma_{\text{DNA}}(z,\phi)=\frac{2\pi\bar{\sigma}}{N}\sum_{j=1}^{N}\int_0^{2\pi}d\phi'\int_{-\infty}^{\infty}dz'\times\nonumber\\
    &\hspace{3.2cm}\psi(z-z',\phi-\phi')\delta(\phi'-\phi_{j}-g(z'-z_{j}))
\end{align}
where $\phi_j$ and $z_j$ describe the coordinates of the strands relative to the defined coordinate system. $N$ is the total number of helical strands ($N=2$ for B-DNA) on the molecule, where each strand is labelled by the index $j$. For DNA, if we define the coordinate system such that $(z=0, \phi=0)$ corresponds to the centre of the minor groove, as in Fig. \ref{fig:Figure 1}, $(z_1,\phi_1)=(0, -\phi_s/2)$, and $(z_2,\phi_2)=(0, \phi_s/2)$, where $\phi_s=0.8\pi$ is the width of the minor groove. We assume that the fixed charges are associated with surface groups centred on helical strands. We account for the finite size of charged groups by introducing the form-factor $\psi(z,\phi)$, normalised as $\int\int \psi(z,\phi)d\phi dz=1$, such that $\psi(q=0,n=0)=1$, where $\tilde{\psi}(q,n)$ is the cylindrical Fourier transform of $\psi(z,\phi)$. Taking the Fourier transform of this expression, we find that for B-DNA:
\begin{align}
   \tilde{\sigma}_{\text{DNA}}(q,n)=4\pi^2\bar{\sigma}\tilde{\psi}(q,n)\delta(q+ng)\cos\left(\frac{n\phi_s}{2}\right)
\end{align}
In the case of thin line charges, $\psi(z,\phi)=\delta(z)\delta(\phi)$, and hence $\tilde{\psi}(q,n)=1$. For the simplest fluctuation case of Gaussian disorder in both $\phi$ and $z$, we can write:
\begin{align}
    \psi(z,\phi)=\frac{1}{2\pi\delta z \delta\phi \erf\left[\frac{\pi}{\sqrt{2}\delta\phi}\right]}&\exp\left[-\frac{z^2}{2\delta z^2}\right]\exp\left[-\frac{\phi^2}{2\delta \phi^2}\right]
\end{align}
where $\delta z$ and $\delta\phi$ are the effective half-width of the distributions, related to the atomic form-factors of the charged groups and the mean-square amplitude of their fluctuations around the `helical lines'. Taking the Fourier transform, in the limit where $\delta\phi\ll\pi$ we obtain:
\begin{align}
    \tilde{\psi}(q,n)=\exp\left[-\frac{1}{2}q^2\delta z^2\right]
    \exp\left[-\frac{1}{2}n^2\delta\phi^2\right].
\end{align}
Considering that in Eq. (27) this expression enters in a product with the Dirac delta function, $\delta(q+ng)$, for $\tilde{\psi}(q,n)$ we can use a simpler expression,
\begin{align}
    \Tilde{\psi}(q,n) = \exp[-\frac{1}{2}n^2g^2\Delta_{\text{eff}}^2],
\end{align}
where 
\begin{align}
    \Delta_{\text{eff}} = \sqrt{\delta z^2+\frac{\delta\phi^2}{g^2}}
\end{align}
is an `effective' half-width of the distribution. Note that Gaussian on- or off- strand fluctuations of the groups around their regular positions on the strands, static or dynamic, result in similar form factors. The incorporation of these form-factors into the theory may further cover the effects due to the finite size of the groups, quenched Gaussian disorder and/or Debye-Waller factors. At the end of this section, we will show how these effects in the radial direction may be accounted for in the simplest way. But here we start first with the case of ‘on-the-surface’ smeared form-factors, described by Eq.(30).  

\subsubsection{Condensed Counterion Distributions}

Highly charged helical molecules, such as DNA, cause adsorption (condensation) of counterions onto their surfaces. The adsorbed counterions are typically more mobile than the fixed surface charges described above, and they may either surround the fixed charges or bind into grooves between the strands formed by fixed charges. We therefore approximate the surface density of adsorbed charges by the inhomogeneously smeared charge density $\sigma_c(z,\phi)$, which follows the same basic helical symmetry as the charged strands. The most general expression we can write for the charge density which satisfies this symmetry is:
\begin{align}
    \sigma_c(z,\phi)=2\pi\bar{\sigma}_c \int_{-\infty}^{\infty}dz'\cdot p(z-z')\delta(\phi-gz')
\end{align}
where we have defined the coordinate system in the same way as above, with $(z=0, \phi=0)$ corresponding to the centre of the minor groove. Here, the subscript $c$ labels parameters for counterions, and $\Bar{\sigma}_c$ is their average surface charge density. We can relate this to the average surface charge density of the DNA $\bar{\sigma}$ through $\Bar{\sigma}_c=-\Theta \bar{\sigma}$, where $\Theta$ is the degree of the overall charge compensation by condensed counterions. $p(z)$ is the probability density of counterion adsorption at the axial distance $z$ from the centre of the minor groove, normalised such that its Fourier transform at $q=0$ is $\tilde{p}(q=0)=\int p(z) dz = 1$. For example, setting $p(z)=\delta(z)$ corresponds to the counterions sitting exactly in the middle of the minor groove. Taking the Fourier transform of Eq.(31), we find:
\begin{align}
    \Bar{\sigma}_c(q,n)=4\pi^2\bar{\sigma}_c\delta(q+ng)\Tilde{p}(q).
\end{align}
Here we can analyse a specific 4-state counterion adsorption pattern, where the smeared probability density is written as:
\begin{align}
    &p(z) = \underbrace{\frac{f_1}{\sqrt{2\pi}\delta z_{c1}}e^{-\frac{z^2}{2(\delta z_{c1})^2}}}_{\text{in the minor groove}} + \underbrace{\frac{f_2}{\sqrt{2\pi}\delta z_{c2}}e^{-\frac{(z-H/2)^2}{2(\delta z_{c2})^2}}}_{\text{in the major groove}}\nonumber \\
    &\hspace{0.8cm}+ \underbrace{\frac{f_3}{2\sqrt{2\pi}\delta z_{c3}}\left\{e^{-\frac{(z-H\phi_s/4\pi)^2}{2(\delta z_{c3})^2}}+e^{-\frac{(z+H\phi_s/4\pi)^2}{2(\delta z_{c3})^2}}\right\}}_{\text{on the strands}} + \underbrace{\frac{2\pi}{L}f_4}_{\text{smeared}}
\end{align}
where different $f_i$ denote fractions of counterions adsorbed in different preferential locations, such that $\sum_i f_i=1$. As labelled in Eq.(33), these locations are; near the centre of the minor groove ($f_1$), near the centre of the major groove ($f_2$), near the charged strands ($f_3$), and randomly distributed along the cylinder surface ($f_4$). Here, we assume that the distribution of ions around each preferential adsorption location is Gaussian, with half-width $\delta z_i$. Despite writing the expression for all possible sites in Eq. (34), for simplicity, in the model below we will only consider ions condensed in the minor and major grooves. Calculating the necessary Fourier transforms, we find that:
\begin{align}
    \tilde{\sigma}^c(q,n)=4\pi^2\Bar{\sigma}_c &\delta(q+ng)\nonumber\\
    &\times\bigg(f_1 e^{-\frac{1}{2}n^2g^2\delta z_{c1}^2} +f_2 (-1)^n e^{-\frac{1}{2}n^2g^2\delta z_{c2}^2}\bigg)
\end{align}

\subsubsection{Bound Charge Distribution} Having described the response of water molecules only through the solvent's nonlocal dielectric function, $\varepsilon(k)$, we do not take into account water molecules specifically adsorbed (bound) to DNA, particularly in the grooves. As mentioned in the introduction, there exists a chiral `spine of hydration' that sits in the minor groove. The water molecules that contribute to this spine form a helical superstructure, with their hydrogens pointing towards the central axis. This leads to a line of non-zero polarisation density we must into account through an additional contribution to $\tilde{\sigma}(q,n)$. 

Indeed, in terms of the volume charge densities, the total charge density is $\rho=\rho_f+\rho_b$, where $\rho_f(\mathbf{r}) = \sigma_{\text{DNA}}(z,\phi)\delta(r-a_{\text{DNA}}) + \sigma^C(z, \phi)\delta(r-a_{\text{CC}})$ is the \textit{free} charge density associated with the fixed DNA and counterion charges, and $\rho_b$ is the \textit{bound} charge associated with specifically adsorbed dipoles. Basic electrostatic identities however reveal a different relationship between the volume and surface bound charge densities 
\begin{align}
    \rho_b(\mathbf{r}) = -\sigma_b(z,\phi)\delta(r-a_w)
\end{align}
where $a_w$ is the radius of the cylinder along which the bound surface charge distribution sits. We can relate this easily to the polarisation density distribution, given that $\sigma_b = \mathbf{P}\cdot\hat{\mathbf{n}}$~\cite{JACK1999}. For a cylindrical surface, we see that $\sigma_b(z,\phi) \equiv P_{\perp}(z,\phi)$ where $P_{\perp}(z,\phi)$ is the radial (normal) component of the polarisation density. Hence, as above, we can write the bound charge distribution of the water spine in the grooves as
\begin{align}
    \tilde{\sigma}_b(q,n)=4\pi^2\bar{P}_0 &\delta(q+ng)\nonumber\\
    &\hspace{-0.5cm}\times\left(w_1 e^{-\frac{1}{2}n^2g^2\delta z_{w1}^2}+(-1)^n w_2 e^{-\frac{1}{2}n^2g^2\delta z_{w2}^2}\right)
\end{align}
where $w_1$ and $w_2$ are the relative fractions of the mean radial polarisation density $\bar{P}_0$ associated with water adsorbed in the minor and major grooves respectively. $\delta z_{w1}$ and $\delta z_{w2}$ are the corresponding widths of their distributions about the centre lines of the grooves. 

Accounting for this contribution, we in a way also account for nonlinear effects of the dielectric about the DNA molecule on the polarisation of water, otherwise considered in the linear response approximation. 
\vspace{0.2cm}
\subsection{Electrostatic potential distribution due to DNA}

Substituting Eqs. (27), (35) and (37) into Eq. (24), we obtain an expression for the electrostatic potential distribution in cylindrical coordinates. Writing the potential as a sum of contributions from the phosphates (DNA), the specifically adsorbed (condensed) counterions (CC), and the structured water (W), the full expression reads:
\begin{widetext}
\begin{align}
    \varphi_{\text{tot}} &= \varphi_{\text{DNA}} + \varphi_{\text{CC}} + \varphi_{\text{W}}\\
    \varphi_{\text{DNA}} &= 8\pi a_{\text{DNA}}\bar{\sigma}\sum_{n=0}^{\infty}\frac{e^{-\frac{1}{2}n^2g^2\Delta^2_{\text{eff}}}}{\delta_{n,0} + 1}\cos[n(\phi-gz)]\cos\left[\frac{n\phi_s}{2}\right]\mathcal{W}_n(R;a_{\text{DNA}},\kappa)\\
    \varphi_{\text{CC}} &= -8\pi a_{\text{CC}}\bar{\sigma}\Theta \sum_{n=0}^{\infty}\frac{f_1 e^{-\frac{1}{2}n^2g^2\delta z_{c1}^2}+f_2(-1)^n e^{-\frac{1}{2}n^2g^2\delta z^2_{c2}}}{\delta_{n,0}+1}\cos[n(\phi-gz)]\mathcal{W}_n(R;a_{\text{CC}},\kappa)\\
    \varphi_{\text{W}} &= -8\pi a_\text{W}\bar{P}_0\sum_{n=0}^{\infty}\frac{w_1 e^{-\frac{1}{2}n^2g^2\delta z_{w1}^2}+w_2(-1)^n e^{-\frac{1}{2}n^2g^2\delta z^2_{w2}}}{\delta_{n,0}+1}\cos[n(\phi-gz)]\mathcal{W}_n(R;a_\text{W},\kappa)
\end{align}
\end{widetext}
As mentioned, for simplicity, we have only considered counterions condensed in the minor ($f_1$) and major ($f_2$) grooves in Eq. (40). It is clear here that the $n=0$ term corresponds to the potential around a homogeneously charged cylinder, and so, the $n\geq 1$ terms are usually referred to as the `helical harmonics'~\cite{KORN2007}. This general formula above can be applied for any case of specific counterion condensation on the DNA molecule. The electrostatic propagator $\mathcal{W}_n(R;a_i,\kappa)$ describes how the potential varies in the radial direction, and for $\varepsilon(k)$ given by Eqs.(6) and (7), this is calculated as:
\begin{align}
    \mathcal{W}_n(R;a,\kappa)&=\tilde{g}_\kappa K_n(\Tilde{\kappa}_nR)I_n(\Tilde{\kappa}_na)+\gamma \Tilde{g}_{L}K_n(\Tilde{q}^n_{d1}R)I_n(\Tilde{q}^n_{d1}a)\nonumber\\
    &\hspace{-0.5cm}+(1-\gamma)\Re\{\Tilde{g}_o H^{(2)}_{n}[(\tilde{g}_n^R-i\tilde{g}_n^I)R]J_n[(\Tilde{g}_n^R-i\Tilde{g}_n^I)a]\}
\end{align}
for $R>a$, where $I_n(x)$ and $K_n(x)$ are the $n$-th order modified Bessel functions of the first and second kind respectively, $H_n^{(2)}(x)$ is the $n$-th order Hankel function of the second kind. In the case of $R<a$, we simply swap the positions of $R$ and $a$ in the Bessel functions. The analytical calculation of this integral, and the definitions of all the parameters are given in the Appendix. In general, these parameters all depend on the roots of the denominator in the integrand of this integral. In the complex plane of $K$, one root is found at $K=\pm i\sqrt{\kappa^2+n^2g^2}$, and the rest are roots of $\varepsilon(\sqrt{K^2+n^2g^2})$; in the case of the approximation we use in Eq.(7), the latter are functions of the parameters $\Lambda_1$, $\Lambda_2$ and $Q$, which relate to the characteristic correlation lengths in the system. For this approximation the result of the integral is split into three terms, each involving certain characteristic lengths. By simply inspecting which lengths are involved with each term, we can have a deeper understanding of the different contributions to the electrostatics in the system. 

The first term arises from ion correlations in the system. Given the simple linear Debye-H\"uckel approximation used in this derivation, we see it follows a simple quasi-exponential decay law where the Debye length $\kappa^{-1}$ is coupled with the DNA helical pitch. The second term arises from exponential water correlations (Lorentzian behaviour) in the system. This is clear given the presence of $\Lambda_1$ in the decay length, also coupled with the DNA helical pitch. Finally the oscillatory contributions are given by the third term. Given the complex arguments of the Hankel and Bessel functions, it is not possible to express this in terms of commonly known special functions. However, we see that this contribution comes from the complex root of $\varepsilon(k)$, and so the oscillation and decay lengths are related to $\Lambda_2$ and $Q$, again coupled with the DNA helical pitch. This is clear given the presence of the parameters $\Tilde{g}_n^R$ and $\Tilde{g}_n^I$ in the arguments of the Hankel and Bessel functions, and that for $n\to0$, $\Tilde{g}_n^R\to Q$  and $\Tilde{g}_n^I\to 1/\Lambda_2$ (see Appendix A for definitions). 

As well as the electrostatic potential, it is convenient for us to also calculate other electrostatic quantities, such as the electric field, $\mathbf{E}$, and the total charge density in the system, $\varrho$. Having obtained the result above for the total electrostatic potential, it is trivial to calculate these quantities, as $\mathbf{E}=-\vec{\nabla}\varphi$ and $\Delta\varphi=-4\pi\varrho$. Importantly these quantities do not diverge in pure, electrolyte-free water, i.e. $\kappa\to0$. This is not the case for the electrostatic potential; if we consider the simpler case of a homogeneously charged cylinder in a bulk dielectric, the electrostatic potential distribution diverges logarithmically at long distances from its axis. Hence, to study the electrostatics of a more pedagogical pure water case, we must consider the electric field and charge density. Of course, the electric field is also of interest per se, as it, to a high degree, determines the force with which DNA will interact with any charged object.

\vspace{-0.5cm}
\subsection{Radial Smearing}
\vspace{-0.25cm}
In addition to the fluctuations of surface charge distributions in the $z$ and $\phi$ directions, we, as promised, now consider fluctuations in the radial direction. Looking again at our definition for the charge distribution in Eq. (20), one of the main assumptions is that the lines of charge sit on a cylinder at a fixed radius, $a$. Such simplification would be suitable had we not included the overscreening effect in the solvent. However, we saw in our brief study of the Born sphere above, accounting for oscillations can lead to overemphasised resonance effects; it is also important to consider similar smearing of the DNA charge distribution in the radial direaction, as it should further suppress the resulting amplitudes of oscillations in the electrostatic potential.

Allowing for smearing of the distribution in the radial direction, we can generalise the expression for the volume charge density of a given helical line $\nu$ as
\begin{align}
    \rho_\nu(z, \phi, r) = s_\nu\sigma_\nu(z,\phi)\zeta(r-\Bar{a}_\nu, \delta a_{\nu})
\end{align}
where $s_\nu = \pm 1$ indicates the sign of the contribution; for free charge density, $s_\nu = 1$, and for bound charge density, $s_\nu=-1$. 
As above, we can consider the case of Gaussian fluctuations, in which case, $\zeta$ takes the form of a truncated Gaussian distribution, as $r>0$: 
\begin{align}
    \zeta(r-\bar{a}_\nu, \delta a_\nu) = \frac{\sqrt{2}\Theta(r)}{\sqrt{\pi}\delta a_\nu\left(1+\erf\left[\frac{\Bar{a}_\nu}{\sqrt{2}\delta a_\nu}\right]\right)}e^{-\frac{(r-\bar{a}_\nu)^2}{2\delta a_\nu^2}}.
\end{align}
where $\bar{a}_\nu$ is the mean radius of the surface along which the helical line runs, $\delta a_\nu$ is the half-width of the distribution, and $\Theta(x)$ is the Heaviside step function. Following the above presented derivation for the electrostatic potential due to charge distributions on a cylindrical surface, we can write an extension of this result for the case of a radially smeared charge distribution, 
\begin{align}
    \bar{\varphi}(\mathbf{r}) = \sum_\nu \hat{\zeta}_\nu s_\nu\varphi_\nu=\sum_\nu\int_0^\infty da_\nu\hspace{1mm}s_\nu\varphi_\nu(\mathbf{r};a_\nu)\zeta(a_\nu-\bar{a}_\nu, \delta a_\nu)
\end{align}
where we term $\hat{\zeta}$ the `smearing' operator, and summation runs over all helical charge motifs $\nu$ associated with DNA (phosphate charges, adsorbed counterion charges and adsorbed water bound charges), and $\bar{\varphi}$ is the smeared potential. By tuning the half-width of the distributions we smear over, we can more precisely account for effects of thermal and structural fluctuations of the DNA molecule.

%% file: 4-theory_results.tex
\begin{figure*}[t!]
    \centering
    \includegraphics[width=\linewidth]{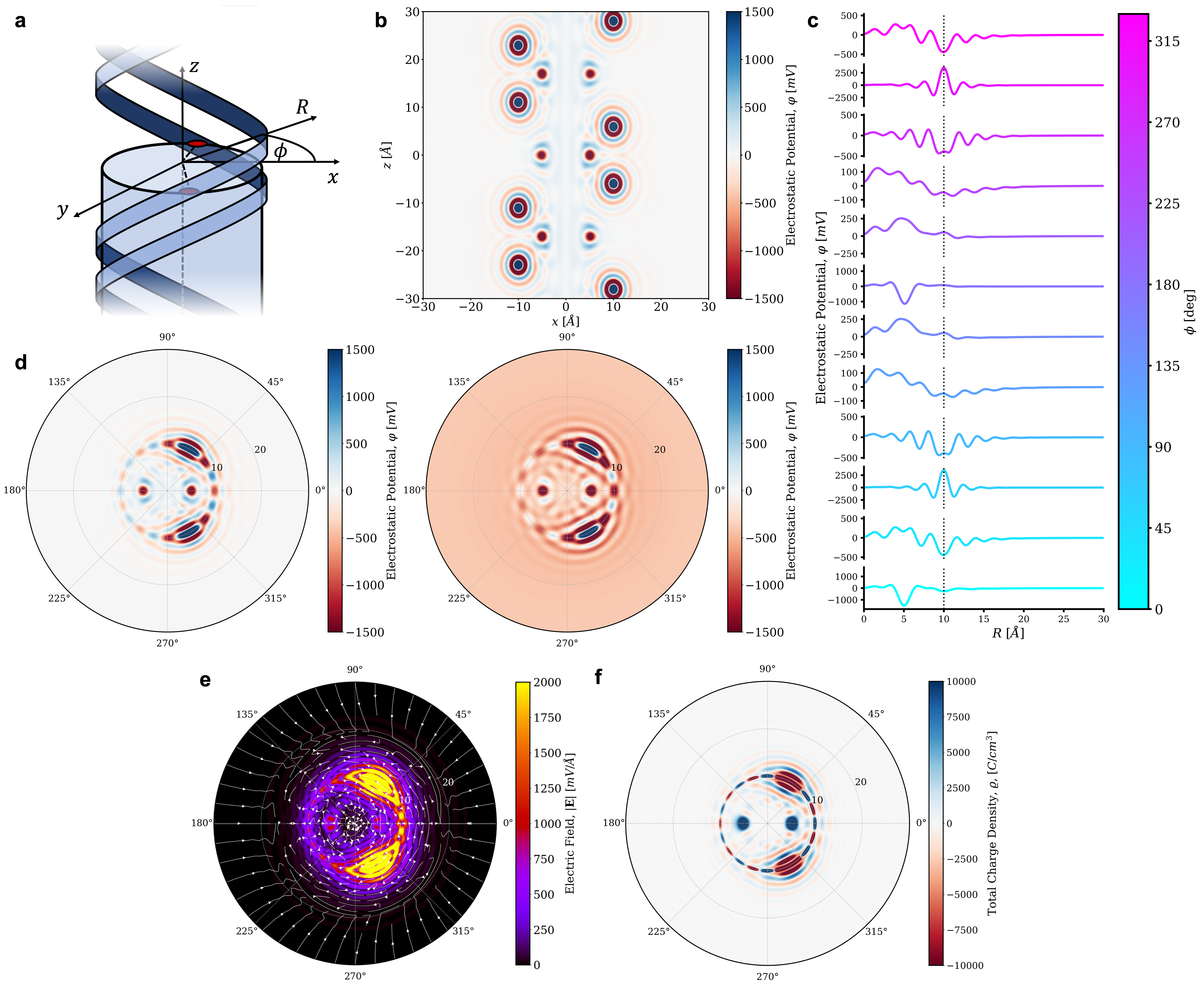}
    \caption{\textbf{Theoretical results for DNA electrostatic potential in solution:} \textbf{(a)} Depiction of system and coordinate axes over which the results are plotted, also displaying the double helical structure of the DNA molecule. For all plots below, the parameters describing the bulk dielectric response of TIP3P are $\varepsilon=94$, $\varepsilon_*=1$, $\Lambda_1=3.67\text{ \AA}$, $\Lambda_2=1.77\text{ \AA}$, $\Lambda_o = 2.13\text{ \AA}$, and $\gamma=0.05$. The parameters describing the DNA geometry and fluctuations are $h_r = 3.4 \text{ \AA}$, $\phi_s = 2.2 \text{ rad}$, $a_{\text{DNA}}=10\text{ \AA}$, $\Delta_{\text{eff}} = 0.25 \text{ \AA}$, and $\delta a_{\text{DNA}} = 0.25\text{\AA}$. For the condensed counter-ion lines, $f_1=0.1$, $f_2=0.9$, $\delta z_{c1} = \delta z_{c2}= 0.75\text{ \AA}$, $\Theta = 0.1$, $a_{CC}= 6\text{ \AA}$ and $\delta a_{CC} = 0.75 \text{ \AA}$. Bound charge distribution parameters are given by $P_0 = -8$ $\mu$Ccm$^{-2}$, $w_1=1$, $w_2=0.75$, $a_W = 5 \text{ \AA}$, $\delta z_{w1}= 0.75 \text{ \AA}$, $\delta z_{w2}=0.75 \text{ \AA}$, $\delta a_W = 0.75 \text{ \AA}$.    
    \textbf{(b)} Electrostatic potential distribution of a DNA molecule in the $xz$ plane, for physiological concentrations ($c_b=0.154$ M, $\kappa = 0.118\text{ \AA}^{-1}$). Phosphate charges located at $x=10 \text{ \AA}$ induce oscillatory behaviour in the electrostatic potential as a result of structured water, whereas condensed counterions and bound water in the grooves of DNA lead to a core of positive potential relative to the bulk. \textbf{(c)} Averaged line-plots of the electrostatic potential distribution as we rotate around the DNA molecule. Each line is averaged over a 30 degree wedge ($\phi\pm 15^\circ$), starting at $\phi=0$. \textbf{(d)} Polar $(R, \phi)$ `cross-sectional' plots at $z=0$ of the electrostatic potential distribution for $c_b = 0.154 \text{ M}$ and at $c_b = 1$ $\mu\text{M}$. \textbf{(e,f)} Magnitude of electric field and total charge density in pure water, i.e. at $\kappa\to0$. Electric field vector directions are drawn as white streamlines in \textbf{(e)}.}
    \label{fig:3-theory}
\end{figure*}

\vspace{-0.25cm}
\section{Theoretical Results \& Predictions}

Plotting Eqs. (38)-(41), applying radial smearing as in Eq. (45), we obtain maps of the electrostatic potential distribution around the DNA molecule in Figure 3. We have plotted the potential in three ways to showcase different behaviours and observations; see Fig. 3(a) to see how they relate to the structure of the double helical molecule. First, in Fig. 3(b), we plot a slice in the $xz$ plane through the centre of the molecule. For the estimated parameters, we clearly see oscillations in the potential propagating from the double helical phosphate lines running about the molecule. As discussed when considering the much simpler Born sphere case, the absolute value of the potential within one molecular diameter of the field source must not be taken literally, as we treat the DNA charge distribution within the `embedded charge approximation' and do not consider the nonlinear response of the medium to the excluded volume of the phosphates. Such approximations therefore lead to the positive and negative `hotspots' at the phosphates and inside the groove, respectively. 

However, while the presence of these oscillations are not a feature to ignore and will be discussed in further detail below, what is particularly interesting here is the presence of a core of positive potential inside the DNA. Such an observation has been previously made in simulation studies of DNA as early as in 1989~\cite{JAYA1989}, but its physical origin is still under debate. Such an effect is also seen at the lipid bilayer/water interface. Some arguments seem to indicate that any positive potential inside non-aqueous systems (like inside the DNA duplex or the lipid bilayer) arises from quadrupolar contributions from water situated at the interface~\cite{HARD2008}. While we do not explicitly consider the quadrupolar moment, it is indirectly accounted for by fitting our model to simulated dielectric response functions. We do show here that such a phenomenon can arise by accounting for adsorbed water by imposing helical lines of polarisation density within the grooves of the DNA molecule. 

When we consider the interactions of DNA in biology, these are often with macromolecules far larger than the size of these oscillations, and so they will often experience a more averaged electric field. The question then becomes whether these interacting molecules will even feel these oscillations. In Fig. 3(c), we plot the potential distribution averaged over 30 degree slices ($\phi \pm 15^\circ$) as we circle the molecule. Even over such a large slice, these oscillations are pronounced close to the minor groove and the phosphate strands. However, these oscillations appear to be weaker in magnitude in the vicinity of the major groove. Such reduced potential in this region compared to the minor groove side can perhaps lead to overall lower electrostatic repulsion, potentially providing a stronger impetus for proteins to approach and specifically bind in this region. Indeed, most DNA-binding proteins will bind to the major groove, where they have better access to the individual nucleotide bases to be `read' by the protein~\cite{PABO1984}. 

Fig. 3(d) showcases the electrostatic potential as a polar $(R, \phi)$ map at both physiological concentration ($0.154$ M) and very low concentration ($1$ $\mu$M). One clear difference when comparing the two is the absence of a positive core at low concentration. This indicates that this effect is not solely reliant on the behaviour of water at the interface. Rather, it is a coupled ion/water effect, where the positive potential can arise from ions screening the phosphate charge from inside the double helix. 

We also see that the oscillatory features remain unchanged between the low and physiological concentration regimes. While such an observation may be clear from examining the form of Eq. (42), this result is not a trivial one, and there is still a lot of debate on the significance of these oscillations. Our recent work on this electrostatic double layer problem with a field-theoretical approach indicates that for millimolar concentrations, these oscillations trap ions, resulting in ion layering at the interface~\cite{HEDL2023}. While the high concentration limit was not studied there, it is clear that ion correlations will take charge, affecting the oscillatory behaviour in the electrostatic potential profile. While these limits have been well studied in varying degrees of complexity, the behaviour of the `intermediate' concentration regime is still unclear. Do physiological concentrations sit within this intermediate regime, and if so, how do these water-water and ion-ion correlation effects interfere or couple with each other? Our results seem to suggest that even at physiological concentrations, the electrostatics are still dominated by the solvent response to the DNA. 

To strengthen this argument, we must make a comparison between physiological and pure water cases to directly observe the effect of electrolyte concentration. However, as we have already noted above, in pure water, i.e. as $\kappa\to0$, it is clear that the potential diverges for the $n=0$ harmonic (i.e. for a homogeneous cylinder). Hence, we can only consider here the electric field and charge density, expressions for which are given in Appendix B, derived from Eqs. (38)-(42). Plotting these electrostatic quantities in Figs. 3(e) and (f) for pure water, we see that these spatial oscillations are purely a consequence of the overscreening dielectric response of water.
\begin{figure}[t!]
    \centering
    \includegraphics[width=\linewidth]{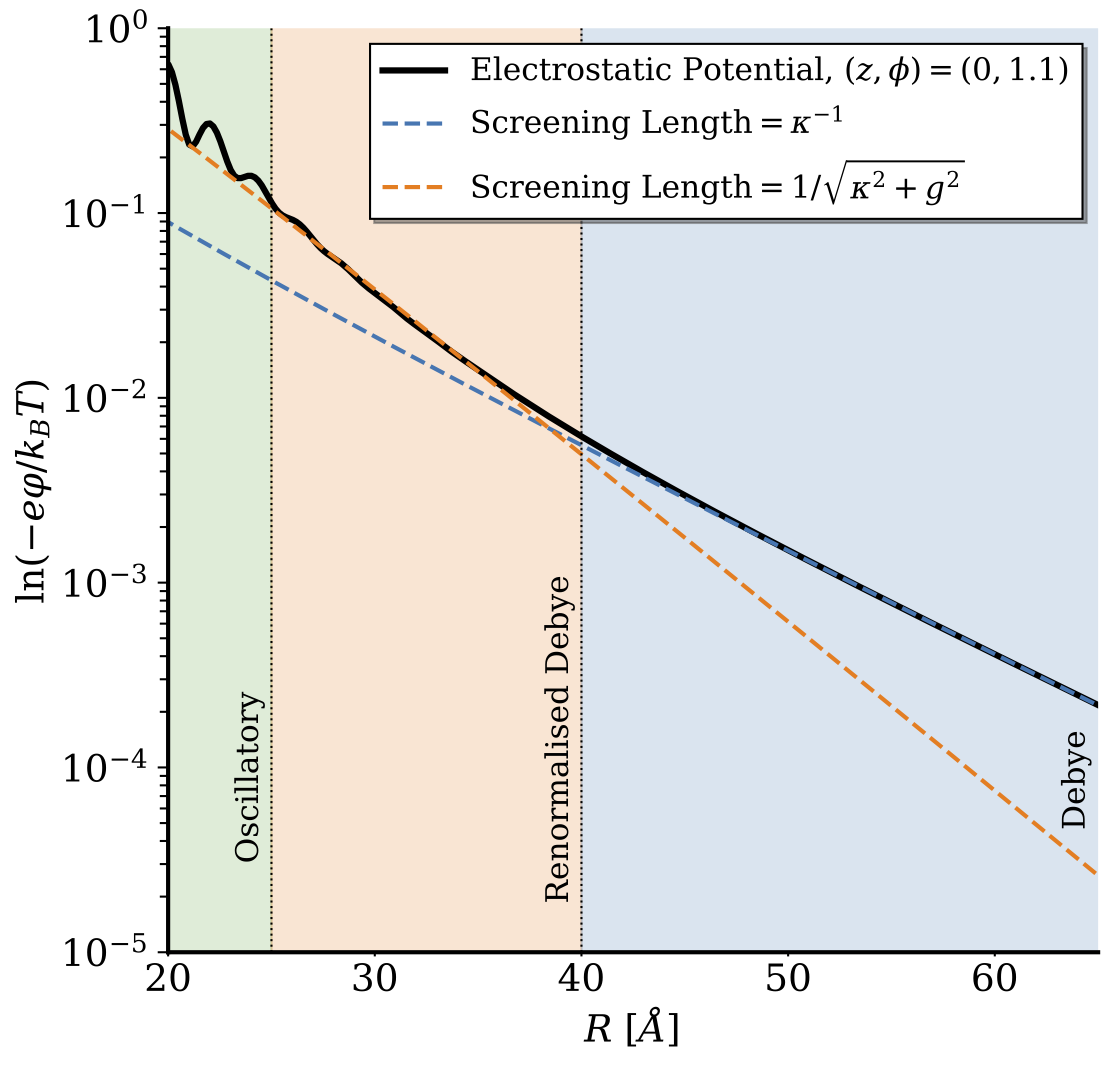}
    \caption{\textbf{Analysis of the long-range Debye-like tail of the radial electrostatic potential distribution of a DNA molecule.} The black curve is plotted for the same parameters as in Fig. 3, at physiological concentrations along $(z,\phi)=(0, 1.1 \text{ rad})$. The blue and orange lines are exponential fits, $f(R)=Ae^{-R/\lambda}/\sqrt{R}$ with decay (screening) lengths $\lambda=\kappa^{-1}$ and $\lambda=1/\sqrt{\kappa^2+g^2}$ respectively. Such fits allow us to identify three screening regions of the DNA molecule: (i) the interfacial 'oscillatory' regime where water dominates, (ii) the renormalised Debye-like regime, where the periodic helicity of the DNA molecule couples with the Debye length, overall reducing the screening length, and (iii) the Debye-like regime, where screening can be approximated by the Debye length of the electrolyte.}
    \label{fig:4-debye-screening}
\end{figure}

Finally, it is of interest to understand the screening of the electrostatic potential, as it determines the electrostatic forces experienced by other charged molecules further away from the DNA. By examining the Debye-like tail of the potential distribution, we can identify three regions as we move away from the DNA molecule; (i) the oscillatory `interfacial' region, (ii) the steeper renormalised Debye tail with screening length $1/\sqrt{\kappa^2+g^2}$ (c.f. Ref.~\cite{KORN1997}), and (iii) the classical Debye tail with screening length $\kappa^{-1}$ (see Fig. 4). This is evident by fitting exponential curves to these different regions, and comparing their gradients against the calculated potential curve. Indeed, such an exponential form would be appropriate to fit the potential as the Bessel functions of the second kind in Eq. (41) are exponentially decaying in their asymptotic large argument limit, such that $\mathcal{K}_n(x)\sim e^{-x}/\sqrt{x}$. 

We summarise the main findings of the theory developed in the first part of this paper below:
\begin{itemize}
    \item At long distances $(R > 40 \text{ \AA})$, we approach the macroscopic limit, and the screening length is simply equivalent to the Debye length, $\kappa^{-1}$.
    \item At intermediate distances, $(25 \text{ \AA} < R < 40 \text{ \AA})$ the double helical structure of DNA coupled to the Debye length, leading to a renormalised screening length of $1/\sqrt{\kappa^2+g^2}$, which is in line with the previous theoretical results based on the primitive model of electrolyte~\cite{KORN1997}.
    \item At short distances $(R<25 \text{ \AA})$ Lorentzian contributions to the dielectric response of water, and consequently a reduced effective dielectric constant close to the DNA surface leads to a dramatic enhancement of the overall electric field.
    \item In this `short-distance' regime, we see strong spatial oscillations in the electrostatic potential. These are a consequence of the overscreening dielectric response of water to the phosphate charges even at physiological electrolyte concentrations, evident from comparing against potential distributions at low concentration, and the electric field and charge density distributions in pure water. The amplitude of these oscillations is reduced the more smeared the helical lines of phosphate and adsorbed counterion charges are. 
    \item Specifically adsorbed water molecules and counterions in the grooves of DNA lead to a positive core of electrostatic potential relative to the bulk.
\end{itemize}

Of course, these results strongly depend on the model of dielectric response of the electrolyte medium used, and so we must either prove or disprove this linear response analysis by comparison with all-atom simulations, the details of which are described in the next section. However, it is obvious that introducing even simple models of nonlocal dielectric response helps to display the important effect of water structure in the electric field created by this `most important molecule'.

%% file: 5-simulation_methods.tex
\vspace{-0.25cm}
\section{Simulation systems \& Methods}

\textbf{Summary of MD simulations - }
We conducted  MD simulations of a 42 bp DNA molecule immersed in an electrolyte solution, varying the composition and the strength of the latter. A typical system~(\fig{Figure 1}b) consisted of~$\sim$1.2~M~atoms and measured 30~nm~$\times$~30~nm~$\times$~14.3~nm. 
The DNA molecule was placed with its helical axis aligned along the $z$-axis and was made effectively infinite by extending the covalent bonds of the DNA backbone over the periodic boundaries of the unit cell. 
The DNA was built to have the structure of a canonical double helix with a 34.28\textdegree~twist per base-pair, which ensured that the ends of the 42~bp fragment perfectly matched at the periodic image boundaries. All non-hydrogen atoms of the DNA were restrained harmonically to their initial coordinates with a spring constant $k_{\rm spring}$, which value was set to 10~\kcalmolAA~(``strong" restraints).

In total, we simulated three systems, differing by the electrolyte conditions. First, we consider the DNA surrounded by a KCl electrolyte of 
physiological, 0.154~M concentration and this system was simulated 
under strong restraints for 125 ns. In a second system, the solvent contained 0.0513~M solution of MgCl$_{2}$, having the same Debye length as the 0.154~M~KCl system under the Debye-H$\mathrm{\ddot{u}}$ckel approximation. This system was simulated for $\sim$500~ns under strong restraints. Finally, we consider a fictitious system whereby charged DNA is considered in pure water, to provide a reference for the realistic systems with electrolyte. This simulation was run for 100~ns under strong restraints.

\textbf{Preparation of the simulation systems - } The 42 base-pair DNA fragment of the poly(AT) sequence was built using 
the Avogadro software~\cite{HANW2012}. The molecule was solvated using the Solvate plugin of VMD~\cite{HUMP1996}. Where needed,
ions were added using the  Autoionize plugin  of VMD  to first neutralize the system and then to produce the desired bulk ion concentration.
The required number of ions was determined from the mass ratio of water and ions, i.e., the  system's molality.

\textbf{Simulation protocols - }
All MD simulations were performed using NAMD2.14~\cite{PHIL2020}, the CHARMM36 parameter set~\cite{HUAN2017A} for protein and DNA, TIP3P water model~\cite{JORG1983},  a custom hexahydrate model for magnesium ions~\cite{YOO2012a} along with the CUFIX corrections to ion-nucleic acid interactions~\cite{YOO2018}. 

Multiple time stepping was used~\cite{BATC2001}: local interactions were computed every 2 fs, whereas long-range interactions were computed every 4 fs. 

All short-range nonbonded interactions were cut off starting at 1 nm and completely cut off by 1.2 nm. Long-range electrostatic interactions were evaluated using the particle-mesh Ewald method~\cite{DARD1993} computed over a 0.11 nm spaced grid. SETTLE~\cite{MIYA1992} and RATTLE82~\cite{ANDE83} algorithms were applied to constrain covalent bonds to hydrogen in water and in non-water molecules, respectively. 

The temperature was maintained at 300 K using a Langevin thermostat with a damping constant of 0.5 ps$^{-1}$, unless specified otherwise. Constant pressure simulations employed a Nose-Hoover Langevin piston with a period and decay of 200 and 50 fs, respectively~\cite{MART1994}. Energy minimization was carried out using the conjugate gradients method~\cite{PAYN92}. Atomic coordinates were recorded every 9.6 picoseconds, unless specified otherwise. Visualization and analysis were performed using VMD~\cite{HUMP1996} and MDAnalysis~\cite{AGAR2011}.

\textbf{Protocols for averaging data over the DNA base pairs - }

To improve the statistical accuracy of our analysis, we averaged the data across frames of the MD trajectories and over 40 base pairs, excluding one at each end of the molecule to mitigate uncertainty arising from wrapping the solvent's coordinates. The analysis per base pair was conducted by choosing cylindrical slabs with a radius of 30~\AA, aligned along the z-axis and partitioned into 40 bins. Each bin had a span of 3.4~\AA~and the base pairs were positioned at the center of each cylindrical bin.
This strictly geometric definition was employed to prevent double counting of atoms caused by their overlap in neighboring bins, thereby ensuring the accuracy of the calculated densities. Successive cylindrical bins were then rotated about the z-axis by 34.28\textdegree, the twist per base-pair for the DNA. 
Subsequently, the coordinates of the transformed atoms were binned on a 900$\times$900 2D lattice.

\textbf{Calculation of the electrostatic potential - } The electrostatic potential was computed using the PMEpot~\cite{AKSI2005} plugin of VMD. 
For each frame of the simulation trajectory, every point charge is approximated by a spherical Gaussian of inverse width $\beta$ (referred to as the ewald factor), normalized to give the original charge upon integration.

The instantaneous distribution of the electrostatic potential corresponding to the instantaneous charge configuration of the frame is obtained by solving the Poisson equation.vThe electrostatic potential maps were obtained by averaging 20–30 ns fragments of the MD trajectories. The instantaneous configurations were then averaged over the MD trajectory, taking frames every 0.25~ns. An ewald factor of $\beta~=~1$~\AA$^{-1}$ and $0.25$~\AA$^{-1}$ was used for the fine and coarse calculations of the potential maps, respectively. The potential maps were stored as a volumetric grid data with a resolution of 0.2~\AA~in the xy plane and $\sim$0.98~\AA~along the z axis. 

To get an average over the DNA base pairs, volumetric slices in z were rotated according to the DNA's twist per base pair and the resolution of the grid in z, $\sim$10.2\textdegree. This was done using the ndimage library in scipy, with a spline interpolation of order 3.

%% file: 6-simulation_results.tex
\begin{figure*}[t!]
    \centering
    \includegraphics[width=\linewidth]{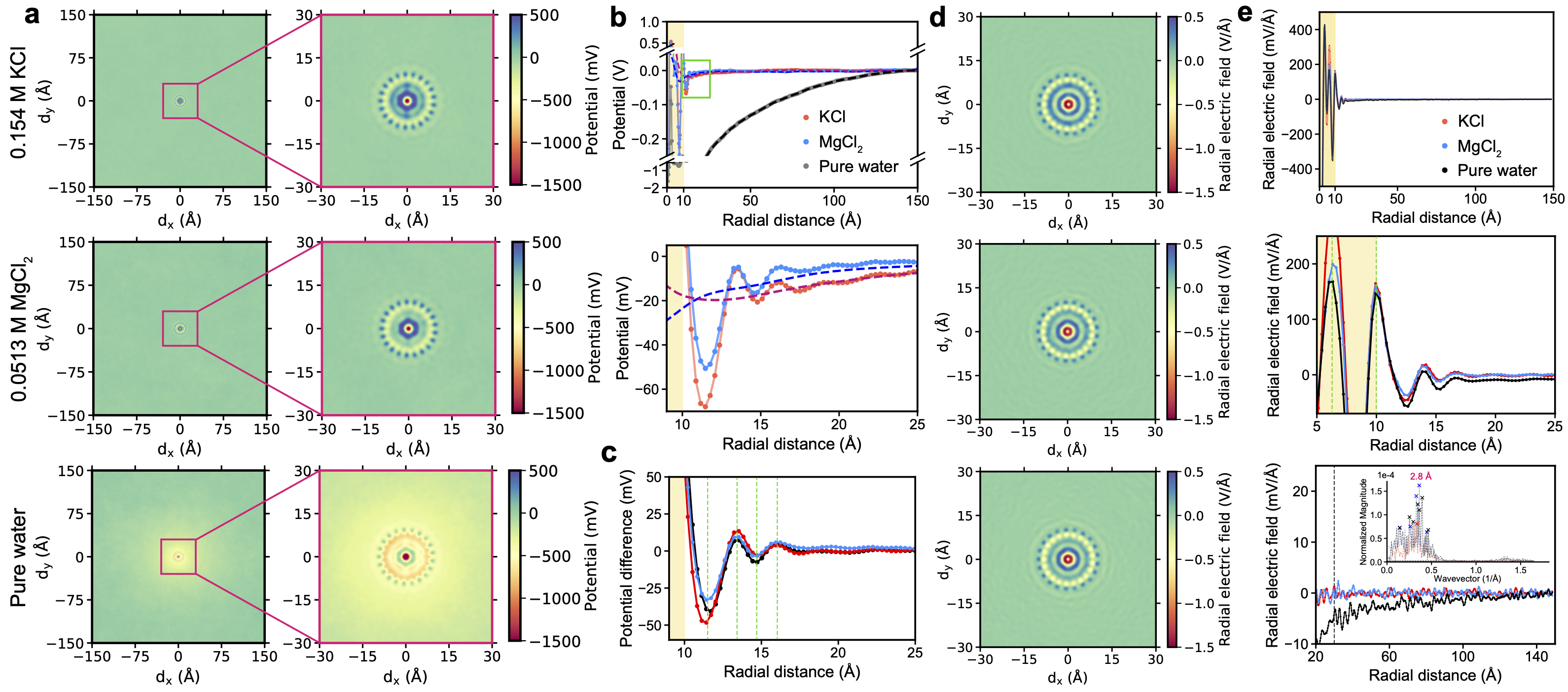}
    \caption{\textbf{Cylindrically averaged electrostatic properties of DNA in solution:} 
    \textbf{(a)} Electrostatic potential maps obtained by averaging the instantaneous distributions of the electrostatic potential over the corresponding MD trajectories and along the z-axis. Data for 0.154 M KCl (top) and pure water (bottom) systems were averaged over the last 100 ns of the corresponding trajectory, sampled every 2~ns. Data for the 0.0513 M MgCl$_2$ system (middle) were averaged over 450 ns and sampled every 4 ns. The electrostatic potential was calculated using the VMD's PMEPot plugin~\cite{AKSI2005} with the Ewald factor of 1~\AA$^{-1}$. \textbf{(b)} Radially averaged profiles of the electrostatic potential for the KCl (red), MgCl$_2$ (blue) and the pure water (black) systems. The bottom plot shows the same data near the DNA, which location is indicated schematically in yellow. Solid and dashed lines correspond to the electrostatic analysis carried out using the fine (Ewald factor of 1~\AA$^{-1}$) and the coarse (Ewald factor of 0.258~\AA$^{-1}$) resolution. \textbf{(c)} The difference of the radially averaged potentials obtained using the fine and the coarse electrostatic calculations. \textbf{(d)} 2D maps of the averaged radial electric field, calculated by locally taking the radial component of the negative gradient of the electrostatic potential map, in cylindrical coordinates. \textbf{(e)} Average profiles of the radial component of the electrostatic field for the three systems. The plots differ by the span of the radial distance. The region occupied by DNA is shown schematically in yellow. Dashed lines (green) indicate the locations of select peaks. The inset (bottom) shows the wavenumber spectra of the averaged electric field difference obtained through the fine and the coarse electrostatic calculations. The wavenumber analysis was restricted to the region 30~\AA\ away from the DNA (dashed black line).}
    \label{fig:5_sim}
\end{figure*}

\section{SIMULATION RESULTS}

\subsection*{Complex electrostatic environment of DNA caused by structured water}

\subsubsection{Cylindrically averaged electrostatics of DNA}

As a baseline for further analysis, we computed the cylindrically averaged electrostatic potential of a DNA molecule (averaged over the $z-$axis of the simulation box) for several electrolyte conditions, similar in spirit to early analytical calculations that assumed a charged cylinder model for DNA~\cite{FRAN1987, MANN1978}. Starting from a fully atomistic MD trajectory, we computed the instantaneous distributions of the electrostatic potential by solving the Poisson equation for each configuration of the partial atomic charges. In doing so, we represented each point charge of a 3D Gaussian density of the mean located at the point charge's coordinates, the inverse width defined by the Ewald factor, i.e., 0.258~\AA$^{-1}$ for coarse and 1~\AA$^{-1}$ for fine resolution calculations, and the total integrated density equal to the point charge value. The instantaneous distributions of the electrostatic potential were averaged over the well-equilibrated parts of the respective MD trajectories and along the $z$-axis (aligned in our coordinate system with the helical axis of the DNA) for the middle 40-base pair section of the helix. The resulting 2D densities are shown in~three panels of \fig{5_sim}a — for the physiological electrolyte condition (0.154~M KCl), a divalent electrolyte of the same screening length (0.0513~M MgCl$_2$), and pure water. 

The presence of ions in the solution produces the expected screening of the electrostatic potential, with the potential decaying away from the DNA much more rapidly in the electrolyte systems in comparison to our fictitious pure water one. The close-up views of the 2D map in the vicinity of the DNA (the right column of \fig{5_sim}a) reveal concentric regions of positive and negative potential common to all three systems. While the presence of such patterns has been expected because the DNA helix itself features a regular pattern of partial charges, the presence of ions is found to profoundly modulate the magnitude and the sign of the potential. Thus, in a background of negative potential due to negatively charged oxygen atoms, the 21 spots of positive potential, located approximately 10~\AA\ away from the helix' centre correspond to the positively charged phosphorous atoms of the DNA backbone, with the number of such spots reflecting the 10.5 base-pair-per-turn periodicity of the helix and the cylindrical averaging of the potential. 

A feature with which we can draw direct parallels with the presented theory is that, in the presence of electrolyte ions, the inner core of DNA is found to bear a positive potential. In pure water, however, this positive core disappears, despite the potential distribution displaying a similar pattern. Guided by the theory, we can deduce that this effect arises from the coupling of structured water dipole and ionic screening in the grooves of DNA, without either of which this phenomenon will not be present. 

Radial profiles of the electrostatic potential in the three systems provide further insight into the effect of ions on DNA electrostatics (\fig{5_sim}b).  The potential at the core of the DNA ($\sim$5~\AA\ away from the centre) is slightly higher in the KCl electrolyte than in MgCl$_2$, which we attribute to the smaller hydration shells of K$^+$ ions which facilitates deeper partitioning of the K$^+$ ions into the grooves of the DNA. 

As expected, the cations of the electrolyte effectively screen the negative charge of DNA backbone, such that the average potential approaches zero already at a distance of $\sim$~30~\AA\ from the helix axis (\fig{5_sim}b top), in contrast to the pure water system, where the potential decays down to the edge of the periodically repeated simulation box ($\sim150$~\AA).

To determine if the effect of water structure on the DNA electrostatics can already be seen in the cylindrically averaged data, we examined the behaviour of the electrostatic potential for the two electrolyte systems in the vicinity of the DNA, i.e.,  
in the region between 9 to 25 \AA\ from the helix axis (\fig{5_sim}b, bottom). 
Gratifyingly, both curves reveal oscillatory modulations of the decaying potential, with the minima and maxima of the
oscillations occurring at similar distances away from the helix. 
To further quantify the oscillating behaviour, we repeated
the electrostatic calculations using a coarser ewald factor of 0.258~\AA$^{-1}$, 
ensuring that the width of the Gaussian approximating each partial charge ($\sim$3.9~\AA) is larger 
than the size of a water molecule and
expecting such a coarse approximation of the atomic charges to wash out the effect of water structure, akin to the smearing effect described above in Section II, and in Refs. ~\cite{KORN1996, HEDL2023}.
Indeed, the potential curves resulting from the low-resolution electrostatic calculations did not exhibit the oscillatory pattern (\fig{5_sim}b, bottom). 
Subtracting the low-resolution profiles from the corresponding high-resolution data isolated
the effect of water structure on the electrostatic potential~(\fig{5_sim}c). 
 We find that the minima and the maxima of the oscillations are indeed located the same distance away 
 from the helix in the KCl and MgCl$_2$ electrolytes and that the distance between the consecutive maxima
 are of the size of a water molecule. 
 Repeating the low-resolution calculations for the third system, DNA in pure water, and subtracting the result from the high-resolution data, we find the oscillations of the potential to also be present in the pure water system, indicating that they arise purely due to the overscreening dielectric response of water to the DNA charge~\cite{BOPP1996, BOPP1998, HEDL2023}. 

Taking the negative gradient of the local electrostatic potential yielded the time-averaged electric field vector at each voxel of each simulation system. The radial component of the electric field was then averaged 
over the z-axis to generate the 2D maps of the electric field (\fig{5_sim}d). 

The resulting maps of the electric field are found to appear similar in all the three systems. 

The average profiles of the radial electric field  elucidate the effect of ions on the local electric field~(\fig{5_sim}e). 
Much like the electrostatic potential, the average radial component of its gradient, the electric field, displays regular alternations; 
however, as components of a vector quantity, these converge to oscillate around zero~(\fig{5_sim}e, top).
In consequence of a slightly higher potential at the core of the DNA ($\sim$5~\AA~away from the centre) in KCl than in MgCl$_2$, the radial component of the electric field is also elevated in the KCl electrolyte, denoted by a dashed (green) line $\sim$6.5~\AA~from the centre~(\Fig{5_sim}e, middle), although the bulk concentrations of the systems being were to have the same Debye length.

Notably, at longer distances, the radial electric field still exhibits some small amplitude but persistent oscillatory patterns~(\fig{5_sim}e, bottom). 
To understand the origin of these oscillations, we analysed the spatial Fourier transform spectra of the longer-range field (R~$>$~30~\AA). 
Similar to electrostatic potential (\fig{5_sim}c), we first subtracted the low-resolution profiles of the radial electric field from the corresponding high-resolution data, thereby isolating the effect of water structure. The Fourier transforms of the resulting radial profiles are shown in ~\fig{5_sim}e bottom, inset.  
All three systems show dominant peaks in the wavenumber spectra around 0.35~\AA$^{-1}$. The corresponding distance in real-space, $\sim$2.8~\AA, is roughly the diameter of a water molecule. It is well known that in molecular simulations of water,  the spatial correlation functions exhibit decaying oscillations which disappear after approximately 1-1.5 nm~\cite{SCIO1995}. Thus the coincidence of the period of these long range oscillations may well be random, simply representing the noise, which can particularly affect the decaying tail. 
\begin{figure}[t!]
    \centering
    \includegraphics[width=\linewidth]{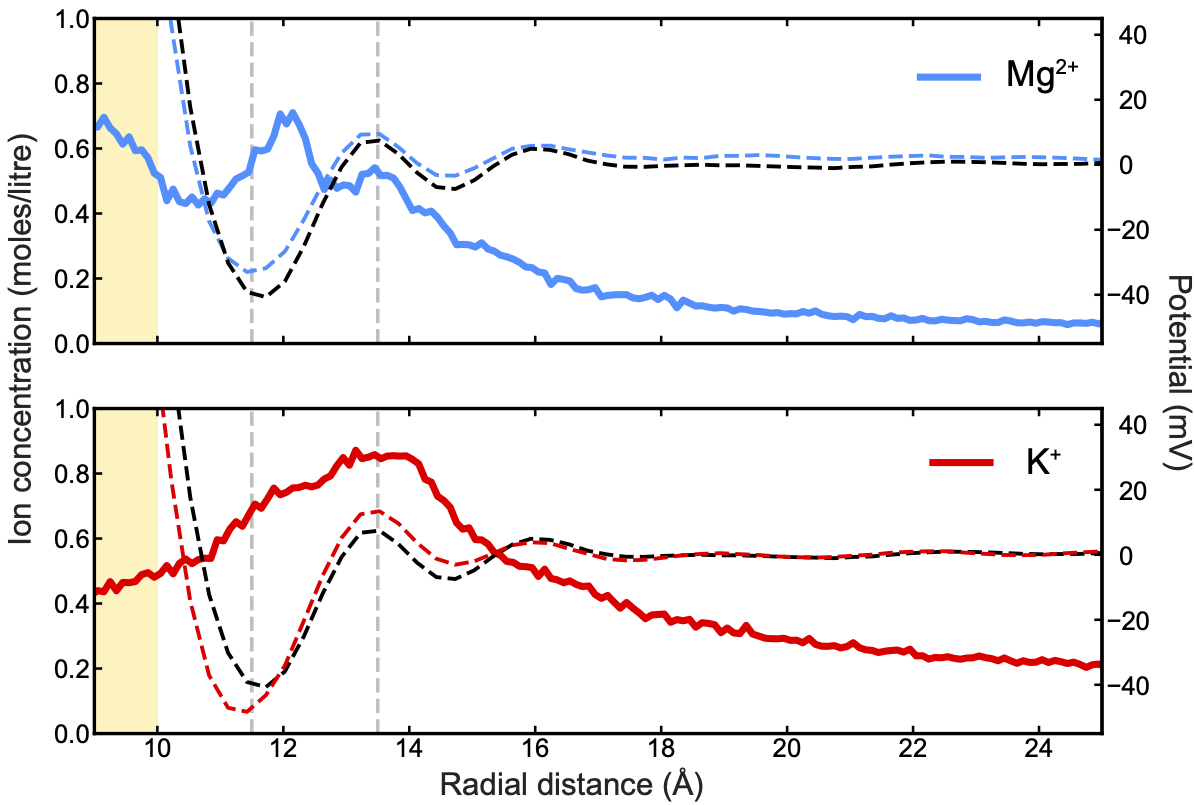}
    \caption{\textbf{Correlation between the cation concentration and the potential difference profiles:} 
    Radial profiles of the cation densities (solid lines), for the Mg$^{2+}$ (top) and K$^{+}$ (bottom) systems respectively, are plotted on the left axis.
    Respective profiles of the electrostatic potential difference induced by structured water are plotted on the right axis, and similarly for pure water is shown by a dashed line (gray). The potential difference profiles are reproduced from \fig{5_sim}c.
    The semi-transparent vertical black lines indicate the locations of select extremum points in the potential difference profiles, at 11.5~\AA and 13.5~\AA~respectively.
    The region occupied by DNA is schematically shown in yellow}
    \label{fig:6_cations}
\end{figure}
 
Bringing our attention back to the consistent large amplitude overscreening oscillations in the short range, the question then comes to how these oscillations influence the interactions of the DNA with charged entities, namely the electrolyte ions. A feature of note in \fig{5_sim}c is that the oscillations of the subtracted potential in the pure water system exactly correlate with the oscillations in the electrolyte systems, which may be explained by preferential localisation of the cations within the wells of the pure water potential. This hypothesis can be tested simply by superimposing the ion density profiles over the corresponding potential distributions (\fig{6_cations}). Remarkably, in the system simulated with Mg$^{2+}$, there is a clear peak in the ion distribution profile at $\sim12$ \AA, corresponding to a well in the potential distribution. This is not unexpected when considering the bulk ion concentration, as the ability for water to layer ions according to its oscillating potential distribution is largely expected in the low ion concentration regime~\cite{HEDL2023}, particularly for the depth of the first well. In the high concentration regime, the ability for water to control the ion distributions is diminished as inter-ion correlations begin to dominate. Physiological concentration sits in some intermediate range between these two regimes, and hence we expected to still observe some signatures of the potential oscillations in the ion distribution profiles in our simulations with K$^+$. Thus, as shown in  \fig{6_cations}, K$^+$ localisation is less pronounced, but there is still some indication in the ion distribution profile corresponding to this effect. In both systems, at larger distances we see that the ions do not obey these oscillating potentials. The reason for this is clear; the depth of the wells quickly become much smaller than $k_B T/e$, and so are not strong enough to fight the entropic urge for ions to spread around.

\begin{figure*}
    \centering
    \includegraphics[width=1\linewidth]{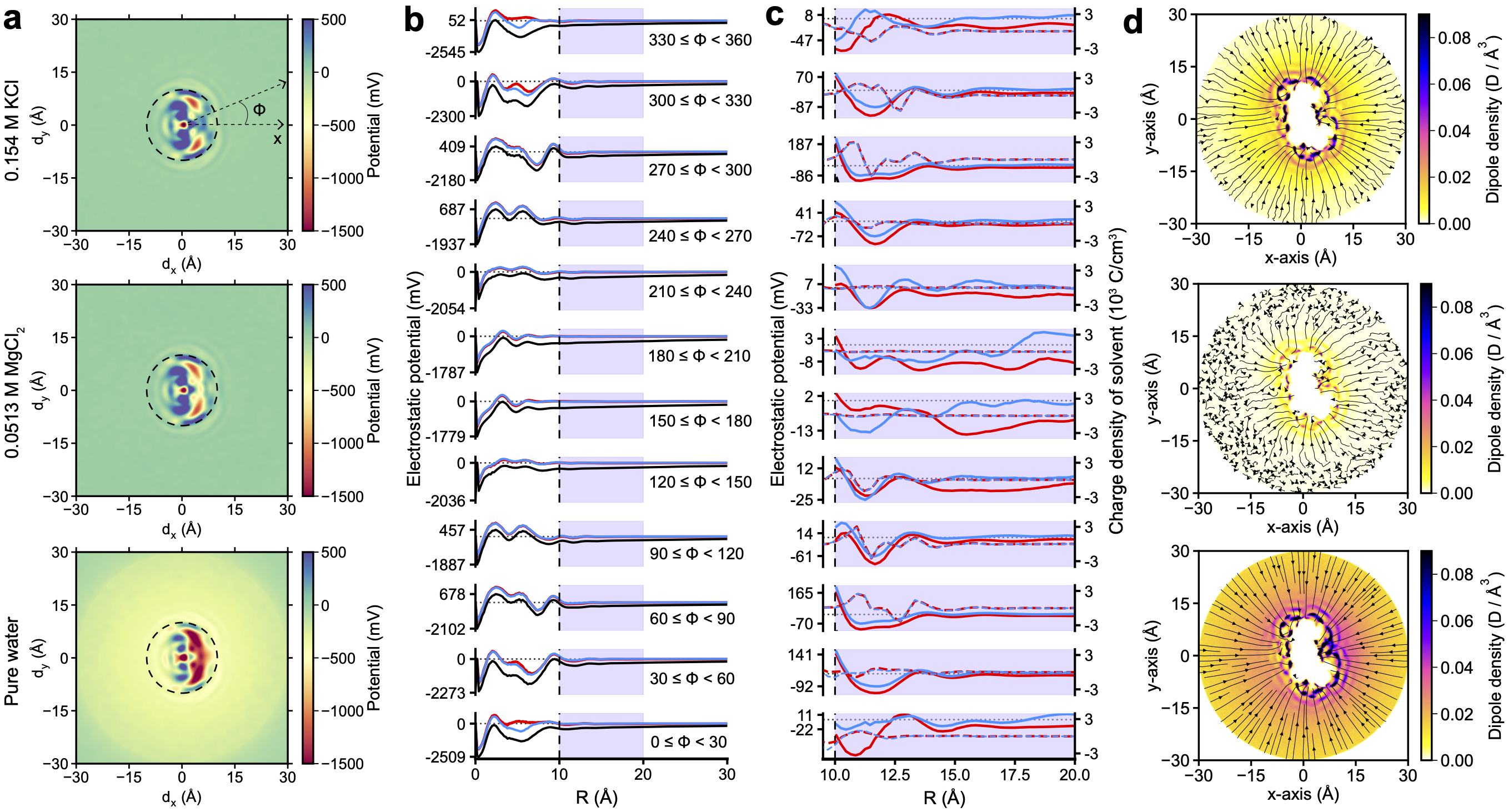}
    \caption{\textbf{Base-pair-reference averaged electrostatic properties of DNA in solution:}
    \textbf{(a)} Projected 2D maps of the electrostatic potential averaged upon alignment of individual base pairs to the reference, for 0.154 M KCl (top), 0.0513 M MgCl$_2$ (middle) and the pure water system (bottom). Excluding the base pair on both ends, the average is based on the last 100 ns for KCl and pure water, and the last 450 ns for MgCl$_2$. The trajectory coordinates were analysed every 2 ns for the KCl and pure water systems or every 4 ns for the MgCl$_2$ one.
    \textbf{(b)} Radial profiles representing consecutive angular patches measured with respect to the x-axis, illustrated by the schematic in the top of panel a. Potential profiles were generated by averaging over voxels within consecutive radial shells of the respective angular patches.
    Line colours follow the same convention as in~\fig{5_sim}, with  K$^{+}$ shown in red, Mg$^{2+}$ in blue and pure water in black.
    The angular range corresponding to each subplot  is indicated below the respective curve.
    \textbf{(c)} Zoomed-in view of the shaded (purple) patch in panel b (10 $<$ R $<$ 20~\AA). 
    The right axis (dashed lines) is the total charge density averaged over the respective angular patches, for K$^{+}$ (red) and Mg$^{2+}$ (blue). The pure water lines are not visible at this scale. Dotted (black) line is a guide to the eye, marking zero for both left and right axes.
    \textbf{(d)} Dipole moment originating from water molecules. Colors in the 2D map represent the local dipole moment density, whereas the streamlines (black) depict the average direction interpolated over the nearby voxels. Any dipole moment arising from separation of ions has been neglected in the analysis of the electrolyte systems. 
    }
    \label{fig:7_sim}
\end{figure*}

\subsubsection{DNA electrostatics in the reference frame of a DNA base pair}

Subsequently, we investigated the average electrostatic properties surrounding the DNA base pair~(\fig{7_sim}). To enhance statistical accuracy, individual slabs of the 3D potential map were aligned with respect to the reference base pair and then averaged (see methods for details). With added electrolyte, the potential profiles for K$^+$ and Mg$^{2+}$ systems decay faster, as expected, with the 2D potential maps displaying striking similarities (\fig{7_sim}a, top and middle).

Notably, these characteristics are also somewhat evident in the fictitious pure water system, albeit, as it should be, with significantly reduced screening levels~(\fig{7_sim}a, bottom). Corresponding line plots compare these systems on a common scale~(\fig{7_sim}b). Inside the DNA core, $0<R<10\text{ \AA}$, the electrostatic potential profiles follow similar trends in all three systems, with a weaker, dielectric screening, in the case of pure water.

In the region just outside the DNA core but in close proximity, $10<R<20\text{ \AA}$, the oscillations are much less pronounced (\fig{7_sim}c).   Oscillations in the presence of electrolyte, K$^+$ (red) and Mg$^{2+}$ (blue) respectively, are well correlated but with subtle variations depending on the azimuthal direction~(\fig{7_sim}c).
The solvent atom charge density is superimposed (\fig{7_sim}b, right axis) with the potential, revealing a close correlation between the local electrostatic potential and the charge density. Minor distinctions arise due to the charge present on the nearby DNA atoms, influencing the local potential while not affecting the solvent charge density. Nonetheless, the overarching patterns exhibit notable correlations.

The uniform characteristics noted in the potential maps across all three systems, irrespective of the presence of electrolyte, prompted us to hypothesise that the preferential alignment of water molecule dipoles may play a pivotal role, and particularly so in the absence of any electrolyte. Having access to atomistic simulation trajectories provides us with a distinctive opportunity to explicitly verify this hypothesis, by tracking the precise locations of water molecules around the DNA. As shown in \fig{7_sim}d, it is evident that water dipoles arrange themselves to minimize the potential energy surface near the DNA molecule. The introduction of cations compensates for the entropic cost linked with the precise localization of water dipoles. A comparison between the pure water system (bottom) and electrolyte conditions (top, middle) reveals higher dipole moment magnitudes, and the interpolated dipole vectors exhibit striking parallel alignment. As expected, the degree of ordering is lower in the presence of Mg$^{2+}$ ions than in K$^+$. In Mg$^{2+}$, the ordering is almost gone around 20~\AA~from the center of the DNA. Note that the dipole moment has only been defined with respect to the water molecules. Smaller contributions arising from localization of ions is neglected in the systems with electrolyte. 

\vspace{-0.2cm}
\subsubsection{Long-range screening} 

The relatively large size of our MD systems allows us to investigate the long-range electrostatic behaviour, namely the Debye decay tail. As mentioned in the theory section, the long-range electrostatic behaviour is important, as it dictates the forces felt by other charged species in solution, which are key to the vast majority of biological processes and interactions~\cite{HONI1995}. It should be noted that examining the long-range behaviour may only be possible in very long simulations of large systems at high ion concentrations~\cite{ZEMA2020}, and may require numerical corrections~\cite{MERE2014}. As our simulations were done at physiological concentrations, we did not expect to find an accurate recapitulations of the theoretical results. Yet, we expected to see distinct screening behaviours enabling direct comparison with the predictions of the theory.
\begin{figure}[t!]
    \centering
    \includegraphics[width=\linewidth]{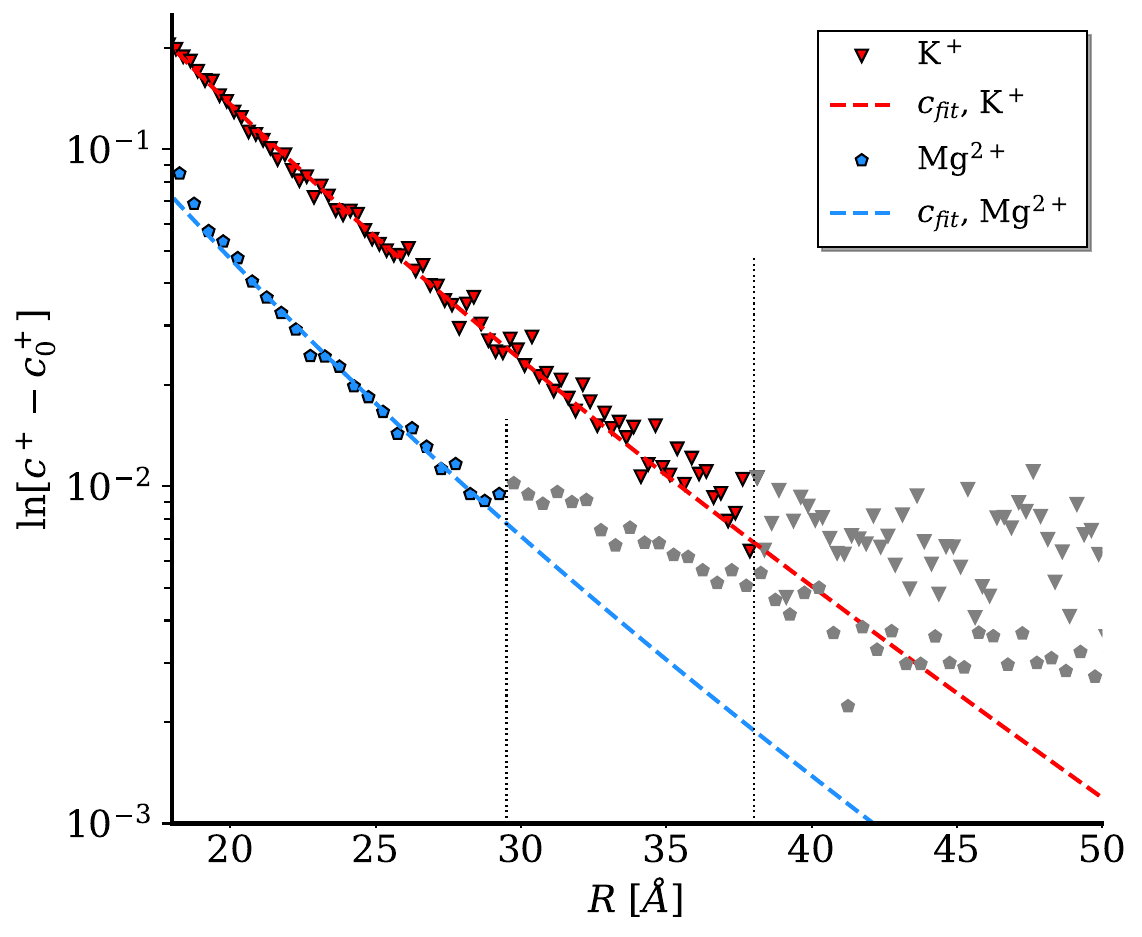}
    \caption{\textbf{Long-range electrostatic screening in MD simulations.}  Radially averaged profiles of K$^+$ and Mg$^{2+}$ concentrations, showing two screening regimes in the electrostatics of DNA. The profiles are fitted using Eq.(46), and the resulting fits are shown as dashed lines. The fit is remarkable accurate in the intermediate range, where the screening behaviour is characterised by the `renormalised Debye length', $\lambda_D = 1/\sqrt{\kappa^2+g^2}$. As reasoned in the text, the reproduction of longer range Debye screening would have required unrealistically large simulation volume and simulation times. The data that might be affected by the finite-size effects (after the transitions between regimes denoted by the vertical dotted lines) is coloured grey. The fitting prefactors are calculated to be; (i) for K$^+$, $A=7.068$, $B=1.528$ and (ii) for Mg$^{2+}$, $A=3.624$ and $B=0.389$.
    }
    \label{fig:8}
\end{figure}

Rather than analysing the electrostatic potential data, we instead look to the ion concentration profiles, which exhibit much less fluctuation. Within the Debye-H\"uckel theory of electrolytes, the concentration profiles, when shifted by the bulk concentration, are proportional to the potential, and so can be analysed in the same way as in the theory. When plotted on a log-scale, both K$^+$ and Mg$^{2+}$ concentration profiles reveal the presence of two screening regimes, Fig.~\ref{fig:8}, transitioning from one regime to another some intermediate distance away from the DNA surface, matching the predictions of the theory. As these ion concentration profiles are roughly proportional to the potential, we can write a simple fitting function inspired by Eqs. (38)-(42),
\begin{align}
    c_{\text{fit}}(R) = AK_1(\tilde{\kappa}_1 R)+BK_0(\kappa R).
\end{align}
To restate, $K_n(x)$ is the $n$-th order modified Bessel function of the second kind, and $\tilde{\kappa}_1 = \sqrt{\kappa^2 + g^2}$, and $g=2\pi/H$, where $H\approx 34\text{ \AA}$ is the helical pitch of the DNA molecule. The prefactors $A$ and $B$ are fitting parameters; within the theory the balance between these will depend on a complex relationship between the counterion adsorption fractions $f_i$ and the overall counterion compensation $\Theta$, which in turn will depend on the specific electrolyte ions we wish to examine. Fitting the data, we focus on the intermediate region between $20<R<40 \text{ \AA}$. Remarkably, we do indeed see the presence of this `renormalised Debye' screening region in both K$^+$ and Mg$^{2+}$ electrolyte simulations, where the Debye length of the electrolyte $\kappa^{-1}$ is renormalised by the helical structure of the DNA to become $\lambda_D\approx 1/\sqrt{\kappa^2+g^2}$. For simulations performed with K$^+$, this screening regime persists until $R\sim 38 \text{ \AA}$, which agrees well with the parameters used for the plots in Figs. 3 and 4. However, this transition occurs at closer distances of $R\sim 29-30\text{ \AA}$ in the Mg$^{2+}$ case, as indicated by the dashed line in Fig.~\ref{fig:8}. 

The quantitative difference in the behaviour of the two electrolytes is understandable when we consider the effect of counterion condensation on the DNA molecule. The location of the transition to renormalized screening length depends on the balance of the fitting prefactors $A$ and $B$, which are functions of the counterion adsorption pattern fractions and the overall charge compensation. Studies examining the adsorption of K$^+$ and Mg$^{2+}$ to DNA suggest that the two ionic species have similar adsorption patterns, both preferentially binding to the major groove~\cite{YOO2012b}, thus leaving the only strong remaining factor in this analysis to be the charge compensation, $\Theta$. Naturally, the divalency of the Mg$^{2+}$ ions will lead to much stronger charge compensation, and hence non-trivial suppression of the different helical harmonics, causing the shift in the cross-over point. 
Note that this behaviour may not be reflected in the fitting parameters because of the inaccuracy associated with modelling the longer range screening behaviour.

It is important to also note that the present linear electrostatic model of the system was not expected to hold for Mg$^{2+}$  at such bulk concentrations. It is well known that the high charge density of the ions leads to strong non-linearity in their electrostatics. However, as a result of our simulations, it seems that also at these physiological concentrations (and hence larger Debye lengths), linear electrostatics can be a valid approximation for biological systems, contrary to the previous arguments~\cite{GROS2002, SHKL1999}.

%% file: 7-conclusions.tex
\section{DISCUSSION \& CONCLUSIONS}
In this work, we have conducted an in-depth joint theoretical and computational study of the electrostatics of DNA. Taking into account all the complexities of the system, namely the double helicity of the DNA molecule, finite size and charge fluctuations through smearing effects, and most importantly for this paper, the nonlocal dielectric properties of the solvent, we have been able to construct a theoretical framework in which these electrostatics can be analysed. 

These theoretical results and predictions have been substantiated by extensive computational characterisation through all-atom molecular dynamics simulations, where we have analysed and observed to high accuracy this complex interplay between the DNA and the solvent and electrolyte ions. In particular, we have been able to understand the following points, consistent through both theoretical and computational analyses:
\begin{itemize}
    \item The electrostatic potential inside the DNA is positive, despite having such a strong negative charge from the phosphate backbone. Such an effect results from the coupled screening effects of structured water in the grooves and electrolyte concentration. 
    \item There are strong electric field oscillations in the near vicinity of the DNA molecule. Analysis of the polarisation density of water shows that these oscillations arise from structured water, both within the grooves of DNA and on the DNA phosphate backbones. 
    \item At physiological concentrations of KCl, the presence of ions does not disrupt these oscillations. This is evident from comparing both the polarisation density (Fig. 7d) and the electrostatic potential profiles (Figs. 5b, 7b) in both pure water and physiological electrolyte. Rather, the ions simply screen the potential, shifting the profile in the positive direction, and only slightly affect the structuring of the water dipoles. This result for the electrostatic potential is also observed for MgCl$_2$, although at the concentration considered, the cations do disrupt this water structuring (Fig 7d, middle) more strongly. 
    \item The electric field distribution across simulations shows little to no difference in the presence or absence of cations, a result that is mirrored by the theoretical model. Such a consistent result across simulations implies that in the close range, water dominates the electrostatics of a DNA molecule at physiological concentrations, validating the results of the theory.
    \item We can see that the first layer of ions in the double layer prefer to be localised in the first potential well created by the overscreening dielectric response of water to the DNA charge. This is the same as to say that these ions physisorb preferentially with their first hydration shell. Further away from the DNA, these trends are less obvious due to the diminished depth of the wells.
    \item At more intermediate distances, outside of the range of water correlations, the Debye length is renormalised, i.e. effectively decreased, by coupling to the double helical structure of the DNA molecule. 
    \item The Lorentzian contribution to the dielectric response of water, which effectively reduces the dielectric constant close to the DNA surface, dramatically enhancing the electrostatic potential and the electric field.
\end{itemize}
Despite the strong comparison between the simulations and the theory, it is important to make note of the number of approximations used within the theory. Firstly, regarding the charge distribution; we do not account for the excluded volume of the DNA itself, as per the `embedded charge approximation' (see Section II.B.1.). Hence, within one molecular diameter $\sim 2.5\text{ \AA}$ around each helical line, the results must be taken with a pinch of salt. This approximation manifests as the strong inversion in the sign of the potential at these points. While smearing of the charge distribution will reduce the effect of this artefact, it is still a strong feature in the theoretical model, and needs to be considered when interpreting its results. Secondly, the approximation used for the dielectric function of the electrolyte is the simplest `interpolation' form we can use. Such a form for the dielectric function still allows analytical calculation of the electrostatic potential, while providing reasonable results. However, it must be acknowledged that the quantitative results we obtain will still depend strongly on the model used. With all this in mind however, given the impressively good agreement with simulations, we can be confident that this analytical, linear nonlocal electrostatic theory captures the key features of the electric field around DNA, and hence emphasises the importance of the solvent structural effects in the electrostatics of this so-called `most important molecule of life'.

%% file: 8-Appendices.tex
\appendix
\section{Electrostatic Propagator}

Eq.(42) contains a number of parameters that were too cumbersome to include in the main text. Their expressions come as a result of taking the integral over $K$ in Eq.(24) by contour integration, with dielectric function defined as in Eqs. (6) and (7). We present these expressions here, in terms of the quantities defined in the main text.
\begin{align}
    \Tilde{\kappa}_n=\sqrt{\kappa^2+n^2g^2}
\end{align}
\begin{align}
    \tilde{q}_{d1}^n = \sqrt{\frac{1}{\Lambda_1^2}+n^2g^2}
\end{align}
\begin{align}
    \tilde{g}_n^R = \sqrt{\frac{Q^2-q_{d2}^2 - n^2g^2 + \sqrt{(Q^2-q_{d2}^2-n^2g^2)^2 + 4Q^2q_{d2}^2}}{2}}
\end{align}

\begin{align}
    \tilde{g}_n^I = \sqrt{\frac{q_{d2}^2 -Q^2+ n^2g^2 + \sqrt{(Q^2-q_{d2}^2-n^2g^2)^2 + 4Q^2q_{d2}^2}}{2}}
\end{align}
where we can clearly see here that $\tilde{g}_n^R\to Q$ and $\tilde{g}_n^I\to q_{d2}= 1/ \Lambda_2$ for $n\to 0$, as stated in the main text. The prefactors for each term in Eq.(42) are given by:
\begin{align}
    \tilde{g}_{\kappa}&=\frac{1}{\varepsilon_*}-\frac{\gamma}{1-\kappa^2\Lambda_1^2}\left(\frac{1}{\varepsilon_*}-\frac{1}{\varepsilon_b}\right)\nonumber\\
    &\hspace{1cm}-\frac{(1-\gamma)(q_{d2}^2+Q^2)^2}{Q^4+2Q^2(q_{d2}^2+\kappa^2)+(q_{d2}^2-\kappa^2)^2}\left(\frac{1}{\varepsilon_*}-\frac{1}{\varepsilon_b}\right)
\end{align}
\begin{align}
    \tilde{g}_L = \frac{1}{1-\kappa^2\Lambda_1^2}\left(\frac{1}{\varepsilon_*}-\frac{1}{\varepsilon_b}\right)
\end{align}
\begin{align}
    \tilde{g}_O = \frac{\pi}{4}\frac{\left(q_{d2}^2+Q^2\right)^2}{Qq_{d2}\left(Q^2-q_{d2}^2 - 2iQq_{d2}+\kappa^2\right)}\left(\frac{1}{\varepsilon_*}-\frac{1}{\varepsilon_b}\right)
\end{align}
\vspace{0.2cm}
\section{Analytical Expressions for Electric Field and Charge Density}
We present here analytical expressions for the electric field, $\mathbf{E}$, and the total charge density, $\varrho$, derived from the electrostatic potential. In cylindrical coordinates,
\begin{align}
    \mathbf{E} = -\left\{\pdv{}{R}\mathbf{u}_R + \frac{1}{R}\pdv{}{\phi}\mathbf{u}_{\phi} + \pdv{}{z}\mathbf{u}_z\right\}\varphi.
\end{align}
$\mathbf{u}_R$, $\mathbf{u}_\phi$ and $\mathbf{u}_z$ are the unit vectors of the three components of the electric field. Each of these components can be split into three contributions due to — (DNA) phosphate charges of DNA, (CC) condensed counterions and (W) the surface bound charge.
\begin{widetext}
    \begin{align}
        E_R^{\text{DNA}} &= -8\pi a_{\text{DNA}}\Bar{\sigma}\sum_{n=0}^{\infty}\frac{e^{-\frac{1}{2}n^2g^2\Delta^2_{\text{eff}}}}{\delta_{n,0}+1}\cos\left[n(\phi-gz)\right]\cos\left[\frac{n\phi_s}{2}\right]\mathcal{W}'_n(R;a_{\text{DNA}}, \kappa),
    \end{align}
    \begin{align}
        E_z^{\text{DNA}} &= -8\pi a_{\text{DNA}}\Bar{\sigma} \sum_{n=1}^{\infty}nge^{-\frac{1}{2}n^2g^2\delta z^2_{\text{str}}}\sin\left[n(\phi-gz)\right]\cos\left[\frac{n\phi_s}{2}\right]\mathcal{W}_n(R;a_{\text{DNA}},\kappa),
    \end{align}
    \begin{align}
        E_\phi^{\text{DNA}}&= \frac{8\pi a_{DNA}\Bar{\sigma}}{R} \sum_{n=1}^{\infty}ne^{-\frac{1}{2}n^2g^2\delta z^2_{\text{str}}}\sin\left[n(\phi-gz)\right]\cos\left[\frac{n\phi_s}{2}\right]\mathcal{W}_n(R;a_{\text{DNA}},\kappa),
    \end{align}
    \begin{align}
        E_R^{\text{CC}} &= 8\pi a_{\text{CC}}\Bar{\sigma}\Theta\sum_{n=0}^{\infty}\frac{\left(f_1 e^{-\frac{1}{2}n^2g^2\delta z^2_{c1}}+f_2(-1)^n e^{-\frac{1}{2}n^2g^2\delta z^2_{c2}}\right)}{\delta_{n,0}+1}\cos\left[n(\phi-gz)\right]\mathcal{W}'_n(R;a_{\text{CC}}, \kappa),
    \end{align}
    \begin{align}
        E_z^{\text{CC}} &= 8\pi a_{\text{CC}}\Bar{\sigma}\Theta\sum_{n=1}^{\infty}ng\left(f_1 e^{-\frac{1}{2}n^2g^2\delta z^2_{c1}}+f_2(-1)^n e^{-\frac{1}{2}n^2g^2\delta z^2_{c2}}\right)\sin\left[n(\phi-gz)\right]\mathcal{W}_n(R;a_{\text{CC}},\kappa),
    \end{align}
    \begin{align}
        E_\phi^{\text{CC}}&= -\frac{8\pi a_{\text{CC}}\Bar{\sigma}\Theta}{R}\sum_{n=1}^{\infty} n\left(f_1 e^{-\frac{1}{2}n^2g^2\delta z^2_{c1}}+f_2(-1)^n e^{-\frac{1}{2}n^2g^2\delta z^2_{c2}}\right)\sin\left[n(\phi-gz)\right]\mathcal{W}_n(R;a_{\text{CC}},\kappa),
    \end{align}
        \begin{align}
        E_R^{\text{W}} &= 8\pi a_{\text{W}}\bar{P}_0\sum_{n=0}^{\infty}\frac{\left(w_1 e^{-\frac{1}{2}n^2g^2\delta z^2_{w1}}+w_2(-1)^n e^{-\frac{1}{2}n^2g^2\delta z^2_{w2}}\right)}{\delta_{n,0}+1}\cos\left[n(\phi-gz)\right]\mathcal{W}'_n(R;a_{\text{W}}, \kappa),
    \end{align}
    \begin{align}
        E_z^{\text{W}} &= 8\pi a_{\text{W}}\bar{P}_0 \sum_{n=1}^{\infty}ng\left(w_1 e^{-\frac{1}{2}n^2g^2\delta z^2_{w1}}+w_2(-1)^n e^{-\frac{1}{2}n^2g^2\delta z^2_{w2}}\right)\sin\left[n(\phi-gz)\right]\mathcal{W}_n(R;a_{\text{W}},\kappa),
    \end{align}
    \begin{align}
        E_\phi^{\text{W}}&= -\frac{8\pi a_{\text{W}}\bar{P}_0}{R}\sum_{n=1}^{\infty} n\left(w_1 e^{-\frac{1}{2}n^2g^2\delta z^2_{w1}}+w_2(-1)^n e^{-\frac{1}{2}n^2g^2\delta z^2_{w2}}\right)\sin\left[n(\phi-gz)\right]\mathcal{W}_n(R;a_{\text{W}},\kappa)
    \end{align}
where in Eqs. (B2), (B5) and (B8), $\mathcal{W}'_n = \partial\mathcal{W}_n / \partial R$.

To incorporate the effect of smearing of the charge distributions, each of these contributions will be smeared with the operator $\hat{\zeta}_\nu$ as defined in the main text. Each component of the electric field will then be $\bar{E}^{\text{tot}}_\beta = \sum_\nu \hat{\zeta}_\nu E_\beta^\nu$  (the bar signifies smearing), where $\beta=\{R,\phi,z\}$, and $\nu =$ \{DNA, CC, W\}. Thus, the magnitude of the smeared electric field is
\begin{align}
    |\bar{\mathbf{E}}_{\text{total}}| = \sqrt{\textstyle{\sum_\beta}(\bar{E}^{\text{tot}}_{\beta})^2}.
\end{align}

The total charge density is proportional to the Laplacian of the potential, 
\begin{align}
    \varrho = -\frac{1}{4\pi}\left\{\frac{1}{R}\pdv{}{R}\left(R\pdv{}{R}\right) + \frac{1}{R^2}\pdv[2]{}{\phi} +\pdv[2]{}{z}\right\}\varphi.
\end{align}
For ease of calculation, we can separate out each term of the total charge density, $\varrho=\varrho_{R}+\varrho_{z}+\varrho_{\phi}$. The `$R$-component', $\varrho_R$ is given by
\begin{align}
    \varrho_R = &-2a_{\text{DNA}}\bar{\sigma} \sum_{n=0}^{\infty} \frac{e^{-\frac{1}{2}n^2g^2\Delta^2_{\text{eff}}}}{\delta_{n,0}+1}\cos\left[n(\phi-gz)\right]\cos\left[\frac{n\phi_s}{2}\right]\mathcal{W}^{(2)}_n(R;a_{\text{DNA}},\kappa)\nonumber\\
    &+2a_{CC}\bar{\sigma}\Theta \sum_{n=0}^{\infty}\frac{\left(f_1 e^{-\frac{1}{2}n^2g^2\delta z^2_{c1}}+f_2(-1)^n e^{-\frac{1}{2}n^2g^2\delta z^2_{c2}}\right)}{\delta_{n,0}+1}\cos\left[n(\phi-gz)\right]\mathcal{W}^{(2)}_n(R;a_{\text{CC}}, \kappa)\nonumber\\
    &+ 2a_{\text{W}}\bar{P}_0\sum_{n=0}^{\infty}\frac{\left(w_1 e^{-\frac{1}{2}n^2g^2\delta z^2_{w1}}+w_2(-1)^n e^{-\frac{1}{2}n^2g^2\delta z^2_{w2}}\right)}{\delta_{n,0}+1}\cos\left[n(\phi-gz)\right]\mathcal{W}^{(2)}_n(R;a_{\text{W}}, \kappa),
\end{align}
where $\mathcal{W}^{(2)}_n = (1/R)\partial(R\mathcal{W}'_n)/\partial R$.  When writing the smeared charge density, $\bar{\varrho}$, it is also convenient to separate $\mathcal{W}^{(2)}_n(R; a_\nu, \kappa)$ into three terms:
\begin{align}
    \mathcal{W}^{(2)}_n = \mathcal{W}^{(2)}_{n, R<a_\nu} + \mathcal{W}^{(2)}_{n, R=a_\nu} +\mathcal{W}^{(2)}_{n, R>a_\nu}
\end{align}
Here, $\mathcal{W}_{n, R=a_\nu}^{(2)} = \delta(R-a_\nu)$; using the truncated Gaussian distribution defined in Eq. (44), the smearing of this term gives
\begin{align}
    \hat{\zeta}\mathcal{W}_{n, R=a_\nu}^{(2)} = \sqrt{\frac{2}{\pi}} \frac{\Theta(R)}{\delta a_\nu\left(1+\erf\left[\frac{\Bar{a}_\nu}{\sqrt{2}\delta a_\nu}\right]\right)}\exp\left[-\frac{(R-\bar{a}_\nu)^2}{2\delta a_\nu^2}\right].
\end{align}
The left and right terms when smeared are calculated numerically. We can also write expressions for the `$z-$' and `$\phi-$' components,
\begin{align}
    \varrho_{z} = &2a_{\text{DNA}}\bar{\sigma}\sum_{n=1}^{\infty} n^2g^2 e^{-\frac{1}{2}n^2g^2\Delta^2_{\text{eff}}} \cos\left[n(\phi-gz)\right]\cos\left[\frac{n\phi_s}{2}\right]\mathcal{W}_n(R;a_{\text{DNA}},\kappa)\nonumber\\
    &-2a_{\text{CC}}\bar{\sigma}\Theta \sum_{n=1}^{\infty} n^2g^2\left(f_1 e^{-\frac{1}{2}n^2g^2\delta z^2_{c1}}+f_2(-1)^n e^{-\frac{1}{2}n^2g^2\delta z^2_{c2}}\right)\cos\left[n(\phi-gz)\right]\mathcal{W}_n(R;a_{\text{CC}},\kappa)\nonumber\\
    &-2a_{\text{W}}\bar{P}_0\sum_{n=1}^{\infty}n^2g^2\left(w_1 e^{-\frac{1}{2}n^2g^2\delta z^2_{w1}}+w_2(-1)^n e^{-\frac{1}{2}n^2g^2\delta z^2_{w2}}\right)\cos\left[n(\phi-gz)\right]\mathcal{W}_n(R;a_{\text{W}},\kappa)
\end{align}
\begin{align}
    \varrho_{\phi} = &\frac{2a_{\text{DNA}}\bar{\sigma}}{R^2}\sum_{n=1}^{\infty} n^2 e^{-\frac{1}{2}n^2g^2\Delta^2_{\text{eff}}} \cos\left[n(\phi-gz)\right]\cos\left[\frac{n\phi_s}{2}\right]\mathcal{W}_n(R;a_{\text{DNA}},\kappa)\nonumber\\
    &-\frac{2a_{\text{CC}}\bar{\sigma}\Theta}{R^2}\sum_{n=1}^{\infty} n^2\left(f_1 e^{-\frac{1}{2}n^2g^2\delta z^2_{c1}}+f_2(-1)^n e^{-\frac{1}{2}n^2g^2\delta z^2_{c2}}\right)\cos\left[n(\phi-gz)\right]\mathcal{W}_n(R;a_{\text{CC}},\kappa)\nonumber\\
    &-\frac{2a_{\text{W}}\bar{P}_0}{R^2}\sum_{n=1}^{\infty}n^2\left(w_1 e^{-\frac{1}{2}n^2g^2\delta z^2_{w1}}+w_2(-1)^n e^{-\frac{1}{2}n^2g^2\delta z^2_{w2}}\right)\cos\left[n(\phi-gz)\right]\mathcal{W}_n(R;a_{\text{W}},\kappa)
\end{align}
where smearing is applied in a similar way to above. These expressions for the magnitude of the electric field and the charge density are plotted in Figs. 3e and f.
\end{widetext}